\documentclass[nonatbib, a4paper, 10pt, oneside, 3p]{elsarticle}

\usepackage{amssymb}
\usepackage{amsmath}
\usepackage{url}
\usepackage{booktabs}
\usepackage{multirow}
\usepackage{tikz}
\usetikzlibrary{positioning, arrows.meta}
\usepackage{subcaption}
\usepackage[dvipsnames]{xcolor}
\usepackage{tabularx}
\usepackage{tcolorbox}
\usepackage{ragged2e}
\usepackage{float}
\usepackage{amsmath}
\usepackage{adjustbox}
\usepackage{enumitem}
\usepackage[natbib=true, style=apa]{biblatex}
\tcbuselibrary{breakable, skins}

\definecolor{color2}{RGB}{64,64,64}

\usepackage{soul}

\definecolor{lightgreen}{RGB}{193,225,193}
\sethlcolor{lightgreen} 

\newif\ifshowchanges
\ifshowchanges
    \usepackage[draft,commandnameprefix=ifneeded]{changes}
    
\else 

\fi

\journal{Information Processing \& Management}

\begin{document}

\begin{frontmatter}



\title{Reducing Confirmation Bias on Controversial Issues through LLM-Generated Multi-Persona Debates} 


\author[1]{Li Shi\fnref{fn1}}
\author[1]{Houjiang Liu\fnref{fn1}}
\ead{liu.ho@utexas.edu}
\author[1]{Yian Wong}
\author[1]{Utkarsh Mujumdar}
\author[1]{Dan Zhang}
\author[1,2]{Jacek Gwizdka\fnref{fn2}}
\author[1]{Matthew Lease\fnref{fn2}}

\affiliation[1]{
    organization={School of Computing, University of Texas at Austin},
    addressline={1616 Guadalupe St},
    city={Austin},
    postcode={78701},
    state={Texas},
    country={USA}
}

\affiliation[2]{
    organization={Institute of Applied Computer Science, Łódź University of Technology},
    addressline={Bohdana Stefanowskiego 18},
    city={Łódź},
    postcode={90573},
    country={Poland}
}

\fntext[fn1]{These authors contributed equally to this work.}
\fntext[fn2]{These authors jointly supervised this work.}

\begin{abstract}
Multi-persona debate systems powered by large language models (LLMs) show promise in reducing confirmation bias; however, empirical evidence remains limited. Through a within-subjects experiment, we compared an LLM-generated multi-persona debate system with an opposing-stances retrieval system. We evaluated system effectiveness in mitigating confirmation bias over 312 study trials across 8 controversial topics, using behavioral measures (changes in eye-fixation patterns) and attitudinal measures (shifts between pre/post-intervention beliefs), with topic familiarity and cognitive tendencies as moderators. Our results show that: (i) participants allocated more than half of their visual attention to attitude-inconsistent content in both systems, with no significant between-system differences; (ii) the debate system showed a suggestive but statistically non-significant buffering pattern among participants with stronger cognitive tendencies, and topic familiarity did not significantly moderate its effect on visual attention; and (iii) reading effort and perceived argument quality influenced participants' engagement with attitude-inconsistent content. These findings suggest that multi-persona debate systems can support engagement with diverse viewpoints, while their comparative advantage over opposing-stances retrieval was not established; individual cognitive tendencies remained an important predictor of visual attention. We provide further implications for designing multi-persona debates to better mitigate confirmation bias, focusing on balancing facts and persuasion and on deciding when to explore diverse viewpoints versus offer concrete solutions.
\end{abstract}




\end{frontmatter}
\clearpage


\clearpage

\section{Introduction} \label{sec:intro}


In traditional information-seeking, users guided by pre-existing beliefs tend to seek out opinions, evidence, and other information that reinforce their beliefs \citep{White2014-zx, Sparrow2011-wd, White2013-ts}. Such \textit{confirmation bias} \citep{berthet2021measurement} can narrow perspectives, entrench existing attitudes, and reduce the rigor and diversity of facts considered in society, ultimately reinforcing oversimplified, dichotomous views of complex issues \citep{Tannen1999-my, Mercier2022-qq, Vedejova2022-cq, Hubeny2025-dp}. This tendency is prevalent with both standard search \citep{Azzopardi2021-zr, Suzuki2021-yn} and newer conversational search systems powered by large language models (LLMs) \citep{Wei2025-ei, Spatharioti2025-cj, Tang2026-sk}. Recent research shows that users tend to guide their interactions with such LLM-based systems by seeking information that supports their pre-existing beliefs and express greater agreement with this confirmatory information \citep{Sharma2024-xx, Khan2024-uo}.


A potential avenue to mitigate this effect is using LLMs to simulate personas with diverse viewpoints, particularly in the context of debate-oriented user experiences \citep{Park2025-mx, Liu2025-uz, Zhang2024-ne}, such as multi-persona debate systems. By enabling users to generate sets of personas that debate a topic from multiple viewpoints, recent studies have found that this system can help reduce echo chambers \citep{Zhang2024-ne} and foster critical thinking at both the individual \citep{Park2025-mx, Song2025-of} and group levels \citep{Chiang2024-lf, Lee2025-ey}.


However, two critical gaps remain in evaluating the effectiveness of LLM-generated multi-persona debates for mitigating confirmation bias. First, there is no direct evidence that these systems can outperform established debiasing strategies commonly used in traditional information retrieval systems, such as presenting evidence of opposing viewpoints. Second, most prior system evaluations have relied primarily on subjective measures, such as user ratings and verbal feedback, without robust quantitative metrics that account for human factors influencing biased information seeking. Notably, people exhibit cognitive tendencies to prefer confirmatory information even when no firm beliefs are present. These thinking dispositions were identified in the context of generative AI that directly influences user information-seeking \citep{Lee2025-vu}. Recent studies also found that individual familiarity with a particular information topic affects people's willingness to engage with that information when it is generated by AI \citep{Topsakal2024-mj}. Both factors signal individual differences that affect the degree to which confirmation bias manifests and the effectiveness of interventions used to mitigate it.  

While prior work suggests that multi-persona debate systems can mitigate confirmation bias, this optimistic claim has not been tested directly against an established debiasing baseline. Rather than assuming the substantial advantage of the debate system, our study puts this claim into a stress-test by comparing a multi-persona debate system with a traditional de-biased information retrieval system.
Specifically, compared to an opposing-stances retrieval system (i.e., presenting both pro and con evidence for a controversial topic), we ask:
\begin{enumerate}[label=\roman*.]
    \item Can a multi-persona debate system more effectively reduce confirmatory visual attention (i.e., perceiving content that aligns with user beliefs) and promote belief change?
    \item Can it counteract bias influenced by individuals' topic familiarity and cognitive tendencies?
\end{enumerate}

To investigate these questions, we conduct a comparative, experimental study, where participants engage with both an opposing-stances retrieval system and a multi-persona debate system to explore a range of social topics. Using a mixed-methods approach, we measure: (i) real-time eye-tracking data that captures participant overt visual attention (i.e., the allocation of individual perception focus by gaze behaviors \citep{Hoffman2016-yi}) to attitude-consistent vs -inconsistent content relative to their prior beliefs; (ii) pre- and post-intervention belief changes; and (iii) comparative qualitative evaluations after using both systems. We also account for individual differences in topic familiarity and cognitive tendencies as moderating factors that may affect the above quantitative indicators (i.e., visual attention and belief change).

Our work contributes to the study of confirmation bias in LLM-powered information systems in three key ways. First, prior work has largely assumed that multi-persona debate systems reduce confirmation bias, but this suggested advantage has not been tested against an established debiasing baseline. We present a direct empirical comparison of an LLM-based multi-persona debate system against a traditional opposing-stances retrieval system for mitigating motivated confirmation bias (i.e., cases where individuals have a strong prior stance), grounded in objective behavioral measures of visual attention and attitudinal measures of belief change. Rather than presuming the debate system's superiority, we test whether it holds, when, and for whom multi-persona debate may be beneficial. Second, we advance understanding of how individual differences, specifically cognitive tendencies towards bias and topic familiarity, moderate the effectiveness of debate-style interfaces. These measures provide a more nuanced, user-centered account of when and for whom such systems are beneficial. Third, we provide methodological grounding for studying confirmation bias in LLM-powered information systems, combining eye-tracking, survey measures, and qualitative analysis within a within-subjects design that controls for individual variability. Together, we offer both empirical evidence and design implications for the development of future LLM-based systems to mitigate confirmation bias.



\section{Related Work} \label{sec:related work}

\subsection{Confirmation Bias in LLM-Powered Information Seeking} \label{subsec:biases}

The rise of large language models (LLMs) has significantly transformed how users seek information online. Instead of manually searching and reviewing information and reading articles through links, users can now quickly access content through LLM-generated summaries or natural language conversations. While this shift reduces the cognitive effort required in traditional retrieval-based experiences \citep{Lee2025-vu}, LLM-powered information systems also amplify human cognitive biases by encouraging reliance on personal heuristics for judgment \citep{Wei2025-ei, Stadler2024-eg, Dai2024-ub, Huang2023-bz}. This results in \textit{confirmation bias}. In psychology, confirmation bias refers to individuals' tendency to prefer and engage with information that supports their existing beliefs \citep{Nickerson1998-sl, Edwards1996-ui}. Confirmation bias can further contribute to the formation of filter bubbles and echo chambers \citep{Quattrociocchi2017-ls} wherein like-minded individuals cluster, communicate, and reinforce shared views, contributing to polarization, misinformation, and societal divides \citep{Garrett2017-qz}. 

Since confirmation bias stems from internal cognitive mechanisms and manifests through behavior, it cannot be observed directly and instead be inferred from observable, measurable actions. These behavioral indicators align with different stages of information seeking, including searching, reading, and evaluating. In psychological research, confirmation bias is examined from two perspectives, depending on whether the behavior indicator is strongly related to an individual's existing beliefs \citep{Nickerson1998-sl, Berthet2022-mx}. \citet{Nickerson1998-sl} distinguishes between \textit{motivated} and \textit{unmotivated} forms: the former operates when people have a personal stake in a topic, whereas the latter may occur even when people are indifferent to a controversial topic. When individuals selectively seek information that supports their existing beliefs, this is considered a \textit{behavioral manifestation} of motivated confirmation bias \citep{Baron1995-jg, Mercier2022-qq}. In contrast, when individuals exhibit a general preference for positive or confirmatory information in the absence of prior beliefs, this reflects an underlying \textit{cognitive predisposition} \citep{berthet2021measurement}.

Scholars studying information-seeking have primarily examined the behavioral manifestations of motivated confirmation bias, showing that users tend to preferentially engage with attitude-consistent information at different stages of the search process. At the exposure stage, prior work demonstrates that users strongly favor congruent information on polarized topics \citep{Vedejova2022-cq}, while those with opposing prior beliefs are more likely to avoid reading information that presents different opinions \citep{Suzuki2021-yn}. At the query formulation stage, similar patterns extend to emerging LLM-powered search tools, where users issue more confirmatory queries and exhibit higher agreement with outputs that align with their prior beliefs \citep{Sharma2024-xx}. At the evaluation stage, this pattern also appears among domain experts. Mental health professionals are more inclined to trust and adopt AI-generated medical recommendations when these are consistent with their original diagnoses \citep{Bashkirova2024-yu}. Although these manifestations are often grouped together as confirmatory behavior, they reflect distinct processes, from the selection of which content to encounter, the allocation of attention across content once encountered, to the subsequent revision or entrenchment of beliefs.

When studying confirmation bias in emerging generative AI systems, scholars have recently highlighted two key considerations. The first is the need to evaluate the influence of the cognitive aspect of confirmation bias \citep{Azzopardi2021-zr}. The cognitive aspect of confirmation bias refers to stable individual differences in thinking dispositions, where higher or lower levels reflect differences in an individual's degree of open-mindedness and analytic thinking style \citep{baron2023thinking}. These dispositions can be reliably measured using validated psychometric instruments and capture systematic variation across individuals \citep{West2008-yw, Rassin2008-wu, berthet2021measurement}. The second key factor is how user familiarity influences confirmation bias in the use of LLM-powered systems \citep{Wei2025-ei, Tang2026-sk}. Prior research focused on traditional information retrieval systems has shown that users heavily rely on their personal familiarity with a topic when evaluating the credibility of online information \citep{Kelly2002-sm, Lucassen2013-pa}. 
This tendency has been further amplified in LLM-powered systems. As information is presented more fluently and persuasively, scholars have found that users are more likely to rely on prior knowledge rather than objective evaluation \citep{Wei2025-ei, Tang2026-sk}.

\subsection{Traditional Debiasing Strategies in IR and HCI} \label{subsec:strategies}

Before the rise of LLMs, research in information retrieval (IR) and human-computer interaction (HCI) had explored various strategies to mitigate confirmation bias in information-seeking. A common approach is to surface attitude-inconsistent content to encourage user engagement with diverse perspectives, particularly in response to motivated confirmation bias. 
Prior work has operationalized this strategy through varying degrees of directness in presenting opposing viewpoints. \citet{Schweiger2014-he} use tag clouds to surface opposing viewpoints, while \citet{Huang2012-ev} directly present counterarguments on a separate screen to reduce cognitive interference and support more balanced exposure.


Another approach targets the cognitive dimension of confirmation bias by encouraging users to scrutinize information more deliberately. Prior work employs interventions ranging from subtle prompts for evaluation, to explicit warnings, and playful visualizations designed to foster reflection. \citet{Sude2021-gh} introduce a voting mechanism for search results to encourage user evaluation of content. Additionally, using a more salient intervention, \citet{Rieger2021-sl} add warning labels such as "this result may reinforce your opinion," requiring users to take explicit action before accessing the content. Moreover, \citet{Kim2021-vx} employ a playful tactic by visualizing diverse online opinions through the metaphor of "starry nights." Taken together, these interventions scaffold reflective thinking through different levels of engagement (from subtle prompts to explicit friction), fostering greater awareness of biased information-seeking behavior.

In addition to external interventions, personalization-aware algorithms have been developed to mitigate bias by adapting information exposure based on user preferences and behavioral tendencies. Specifically, \citet{Schwind2012-wg} propose search engines that highlight results inconsistent with inferred user preferences. Moreover, \citet{Jasim2022-rm} introduce a recommendation model that balances consumer reviews associated with different emotional responses. Collectively, these approaches intervene algorithmically through different mechanisms to support more balanced information exposure without requiring deliberate user effort.


\subsection{Multi-Persona LLMs as a New Approach for Simulated Deliberation} \label{subsec:llm simulation}

Multi-persona LLMs are emerging as a new information-seeking paradigm to mitigate confirmation bias, offering persona-driven, conversational, and multi-turn interactions that differ substantially from the source-grounded, static, and evidential presentation of traditional retrieval-based systems.
The core idea is to use LLMs as proxies for human interlocutors, simulating different perspectives and engaging in constructive dialogue to encourage meaningful deliberation \citep{Kambhatla2024-cn, Argyle2023-mz, cho-may-2020-grounding, Tessler2024-gl}.
Researchers create LLM personas through prompt engineering by aligning LLM outputs with diverse human perspectives, goals, and behaviors \citep{Park2022-qp, Prpa2024-dx, Salminen2024-ik}. Personas may embody different identities, ideologies, or lived experiences, enabling LLMs to engage in richer, more contextually grounded, conversation-based social exchanges \citep{Shanahan2023-kz, Chen2024-ru}.


Researchers have suggested that using LLMs to simulate multi-personas for debate could mitigate the confirmation bias resulting from standard LLM-powered systems \citep{Zhang2024-ne, Song2025-of}. Unlike traditional debiasing strategies, the diversity of LLM-generated multi-persona debates can scaffold user critical thinking from a third-person perspective rather than reinforcing biases tied to their own first-person viewpoint. The debate structure can take several forms: (i) binary opposition, where a persona simply counters a user belief or decision to provoke reflection or justification \citep{Chiang2024-lf, Lee2025-ey}; (ii) multi-option deliberation, where several personas advocate for different viewpoints or options in a decision-making scenario, such as shopping \citep{Park2023-ti} or verifying fake news \citep{Liu2025-uz}; or (iii) nuanced roundtable dialogue, where different personas represent differing sociodemographic or ideological standpoints, engaging in conversational exchanges to spur critical thinking \citep{Zhang2024-ne}.

Recent studies have empirically demonstrated that LLM-generated multi-persona debate helped mitigate confirmation bias. \citet{Zhang2024-ne} develop a multi-agent dialogue system for social media discussions. They found that participants' use patterns (e.g., progressively exploring diverse personas) suggested greater openness to diverse viewpoints. Additionally, \citet{Song2025-of} conducted a study in which participants discussed social issues either with a single persona or with several distinct personas. Their results showed that interacting with multiple personas produced larger changes in participants' initial views, but this effect was accompanied by a higher sense of social pressure. 


\subsection{Prior Study Gap and Our Research Focus}
Although emerging studies have promoted using LLM-generated multi-persona debates to reduce confirmation bias, there are remaining significant gaps in evaluating the effectiveness of such systems:
\begin{enumerate}[label=\roman*.]
    \item No direct comparisons have been conducted to determine whether LLM-generated multi-persona debates provide substantive benefits over established debiasing strategies in traditional information retrieval systems for mitigating confirmation bias. Additionally, it is unclear when and for whom multi-persona debate may be beneficial.
    \item Prior studies do not account for the role of individual cognitive dispositions and topic familiarity in influencing the effectiveness of LLM-generated multi-persona debates for mitigating confirmation bias.
\end{enumerate}

Our research fills these gaps by directly comparing a multi-persona debate system to an opposing-stances retrieval-based system (a representative traditional debiasing approach) and incorporating both behavioral and cognitive dimensions of confirmation bias. Specifically, we adopt three complementary approaches. First, from a behavioral perspective, we examine whether the debate system attracts more visual attention (measured using eye-tracking gaze data) and whether this leads to belief change (measured through pre- and post-intervention self-reported belief scores) when participants hold a clear prior stance. Second, from a cognitive perspective, we study how individuals' cognitive tendencies and topic familiarity (measured before the experiment) influence visual attention and belief change. Third, we complement these quantitative analysis with post-hoc qualitative insights from think-aloud protocols and user reflections to better understand the real-world effectiveness of multi-persona debates.

\section{Methods} \label{sec:methods}

We conducted a controlled within-subjects study by comparing a multi-persona LLM-powered debate system with a baseline retrieval system that presents opposing viewpoints. In this section, we describe our formal evaluation protocol, including the study procedure (\ref{subsec:exp_design}), experimental apparatus (\ref{subsec:system}), measurements (\ref{subsec:measurement}), hypothesis (\ref{subsec:hypothesis}), and data analysis (\ref{subsec:analysis}).

\subsection{Study Procedure} \label{subsec:exp_design}

\subsubsection{Tasks} \label{subsec:task design}

Eye-tracking studies typically employ within-subjects study designs. Examples include studies on attentional bias toward food \citep{Hummel2017-fg}, UX researcher perceptions of persona informativeness \citep{Salminen2018-yn}, decision-making efficiency of in-person versus remote collaboration \citep{Wisiecka2023-bq}, risk-related option evaluation \citep{Glockner2012-rg}, and assessments of news factuality \citep{Shi2023-li}. We also adopted this approach to account for individual variability, which is particularly pronounced and sensitive across participants in eye-tracking data \citep{Staub2021-cb}.

\begin{figure*}[ht]
    \centering
    \includegraphics[width=0.9\textwidth]{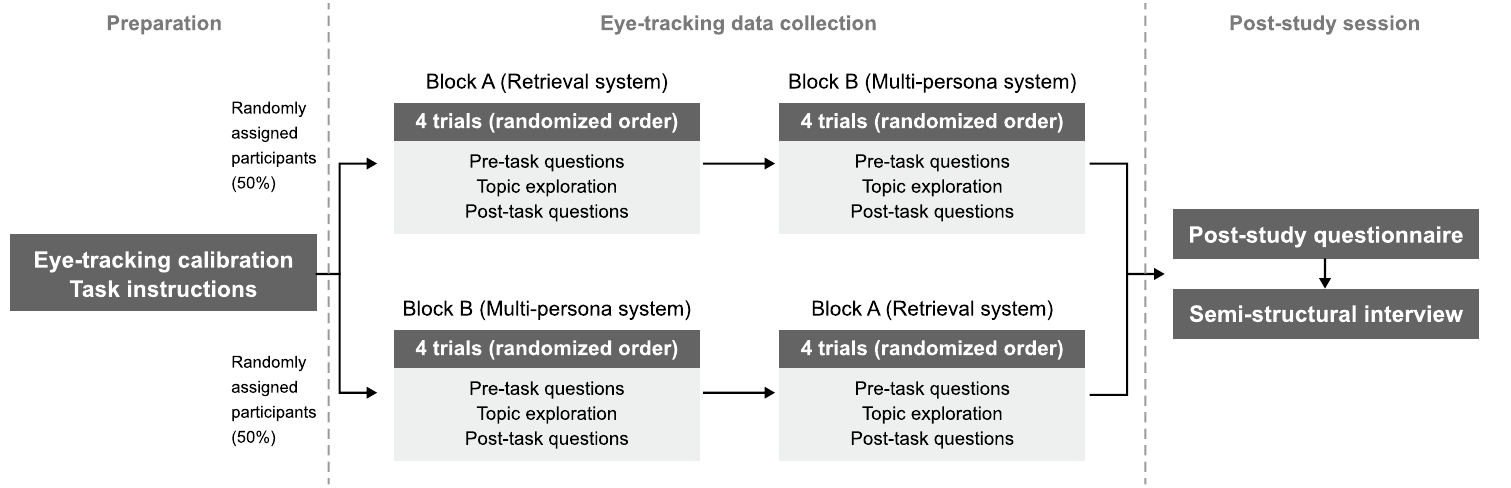}
    \caption{Flowchart of the experimental procedure. Block A (retrieval-based) and Block B (multi-persona debate) are counterbalanced parallel conditions. Participants are randomly assigned to begin with either block, and topic sets are non-overlapping across blocks to prevent content carryover. Absence of significant order effects confirmed via reliability checks (Section \ref{sec:reliability})}
    \label{fig:procedure_diagram}
\end{figure*}

As illustrated in Figure~\ref{fig:procedure_diagram}, the experiment began with eye-tracking calibration followed by task instructions. Participants were then randomly assigned to one of two study blocks (i.e., Block A or Block B) and later switched to the other block to experience both systems. Each block started with a training task to help participants familiarize themselves with the assigned system.

Within each block, participants completed four randomized trials assigned to different topics. Each trial consisted of three parts: pre-task questions, topic exploration, and post-task questions. In the pre-task questions, we asked participants to explicitly rate their topic familiarity, pre-existing belief (i.e., stance) of the topic, and confidence in their stance. After participants completed the trial, they again reported their current stance on the topic and the confidence score (in the post-task questions). Once they completed four trials in one block, they moved on to the other block and used the alternate system. 

When transitioning to a new block, participants took a brief break of approximately 5 minutes, followed by a training session lasting about 10 minutes. The break served as a washout interval to reduce potential order effects. During the training session, participants first used the other system to explore a randomly selected topic that was not included in the main study, allowing them to become familiar with the tool. This procedure ensured that participants were sufficiently comfortable using the system before data collection began and helped minimize potential confounds between system usability and engagement with the content. 

Upon completion of all eight trials across both systems, participants were asked to complete a post-test questionnaire to assess their propensity of cognitive tendencies towards confirmatory information. Following this, a semi-structured interview was conducted to gain insight into user interactions with the systems. 

To minimize order effects, both the order of system exposure (Block A and Block B) and the sequence of topics (trials) were randomized across participants. Task topics were divided into two sets (see Table \ref{tab:topicality}). For each participant, one set of topics was randomly assigned to the first system, with the other set of topics used with the second system. Within each block, the four topics were further randomized in presentation order. This design aimed to reduce potential confounds from topic familiarity, individual cognitive tendencies, and the order of interface exposure that could bias the perceived effectiveness of each system. We also conducted additional reliability checks, which confirmed the absence of significant order effects (reported in Section \ref{sec:reliability}).

Details of the pre- and post-task questionnaires and interview questions are presented in \ref{appendix:measurement}. The entire experimental study took approximately 1.25-1.75 hours to complete per participant: 60-80 minutes for the experimental portion and 15-25 minutes for the interview. 

\subsubsection{Topics} \label{subsec:topic selection}
We selected eight topics from ProCon.org\footnote{ProCon.org: \url{https://www.britannica.com/procon}} to span a diverse range of categories while ensuring that topics were also relevant to our participant pool of university students (Table \ref{tab:topicality}). Topics are formulated as yes/no proposed actions, such as "Should abortion bans be illegal?" or "Should space be colonized?", wherein each topic inherently embeds a stance supporting or opposing the given proposed action.

\begin{table}[ht]
\centering
    \begin{tabular}{@{}lll@{}}
    \toprule
    \textbf{ID} & \textbf{Topics} & \textbf{Area} \\ \midrule
    1 & Should animals be used for scientific or commercial testing? & Science \& Technology \\
    2 & Should humans colonize space? & Science \& Technology \\
    3 & Should vaping with e-cigarettes be legal for minors? & Health \& Medicine \\
    4 & Should abortion be legal? & Health \& Medicine \\
    5 & Should the Federal corporate income tax rate be raised? & Economy \& Taxes \\
    6 & Should the United States implement a universal basic income? & Economy \& Taxes \\
    7 & Should TikTok be banned in school? & Entertainment \& Sports \\
    8 & Do violent video games contribute to youth violence? & Entertainment \& Sports \\
    \bottomrule 
    \end{tabular}
\vspace{0.2cm}
\caption{Topics and their domain areas. Topics span four domains classified by ProCon.org, including Science \& Technology, Health \& Medicine, Economy \& Taxes, and Entertainment \& Sports. We select two topics per domain to ensure balanced coverage.}
\label{tab:topicality}
\end{table}


Our study focuses on motivated confirmation bias, in which participants hold clear prior beliefs about the topics they encounter. In this context, the distinction between attitude-consistent and attitude-inconsistent information underlying our eye-tracking measures is only meaningful when a directional prior belief is present. Accordingly, we selected controversial topics that frequently appear in popular discourse, to ensure that participants would hold clear prior beliefs about them.

Additionally, to balance the limitations of single-topic studies with the need for cross-topic generalizability, we followed prior work in selecting topics across multiple domains \citep{Guess2020-do, Vedejova2022-cq, Wang2024-hg}. Building on this rationale, our topic set spans a wide range of issues, enabling evaluation of whether the multi-persona debate system can foster more open-minded engagement than the opposing-stances retrieval-based system.

\subsubsection{Participants recruitment} \label{subsubsec:recruitment}

We conducted the study in a university usability lab with $N=40$ participants (19 women, 18 men, and 3 non-binary individuals)\footnote{This sample size is based on a priori power analysis we conducted to detect a potential medium effect size with a statistically significant difference. Because the required sample size is strongly influenced by our specific study design and hypothesis formulation, we further elaborate on details in Section \ref{subsec:hypothesis}.}. Participants were recruited using convenience sampling and were aged between 18-34 (demographic details are provided in \ref{appendix:participant demographics}). The participants come from a variety of academic disciplines, ranging from applied sciences, natural sciences, and engineering to social sciences, bringing diverse educational backgrounds. Overall, participants tended to be liberal: 29 (72.5\%) described themselves as strongly liberal, liberal, or moderately liberal. Eight participants (20.0\%) reported being centrists with a mix of liberal and conservative views, and three (7.5\%) identified as moderately conservative. All were pre-screened for native-level English familiarity and 20/20 vision (uncorrected or corrected). Our study was approved by our Institutional Review Board (IRB), and each participant was compensated with a \$30 USD gift card upon study completion.

Prior to the experiment, all participants were informed through the consent process that the debate personas and their arguments were AI-generated by a large language model. No deception regarding the nature of the content was employed. During the task, participants actively constructed and customized the LLM-generated personas while using the debate system.



\subsection{System Apparatus} \label{subsec:system}

\subsubsection{Eye-tracking device}
We used the Tobii TX-300 eye-tracker and processed eye-tracking data through Tobii Pro Lab (Version 24.21)\footnote{\url{https://www.tobii.com/products/software/behavior-research-software/tobii-pro-lab}}, a commercial usability and eye-tracking software that supports raw gaze data collection, user interaction recording, and preliminary data cleaning for subsequent analysis. 

\subsubsection{Baseline opposing-stances retrieval-based system}
We adopted ArgumentSearch\footnote{\url{https://argumentsearch.com/}} \citep{Daxenberger2020-cg} as our baseline system (Figure~\ref{fig:baseline}). ArgumentSearch is a retrieval-based search engine that identifies relevant articles from the open web and organizes content snippets based on their stance toward the user query, either supportive or opposing. We selected it over general-purpose search engines such as Google or LLM-powered conversational tools like ChatGPT, which have been shown to reinforce biased query formulation \citep{Sharma2024-xx}. By using ArgumentSearch, which represents a traditional debiasing strategy, we were able to assess the multi-persona debate system against a realistic and representative point of comparison.

Unlike common design patterns in traditional search systems, such as one-dimensional result ranking or single-agent conversations, ArgumentSearch explicitly categorizes search results by stance, providing a clearer and more structured presentation of comparative information. We selected it not as a bias-neutral baseline but as a representative implementation of a well-established traditional debiasing strategy, allowing us to assess the added value of multi-persona debate over an already debiased retrieval design.

Additionally, to ensure content consistency across participants, all topic queries were pre-specified and identical as listed in Table \ref{tab:topicality}, with retrieval constrained to a fixed time-range window. No participant-generated query variation was permitted, ensuring that the content retrieved was consistent across all participants for each topic.

In the following sections, we refer to this opposing-stances, retrieval-based system as the \textbf{baseline system} for brevity.


\begin{figure}[p]
    \centering
    \includegraphics[width=\textwidth]{Figures/baseline.pdf}
    \caption{A screenshot of the baseline ``ArgumentSearch '' system. When a user enters a query, such as ``Do violent video games contribute to youth violence?'' the system identifies relevant articles and presents content snippets representing both sides, including those in favor (PRO) and those opposed (CON) to the topic.}
    \label{fig:baseline}
\end{figure}

\begin{figure}[p]
    \centering
    \includegraphics[width=\textwidth]{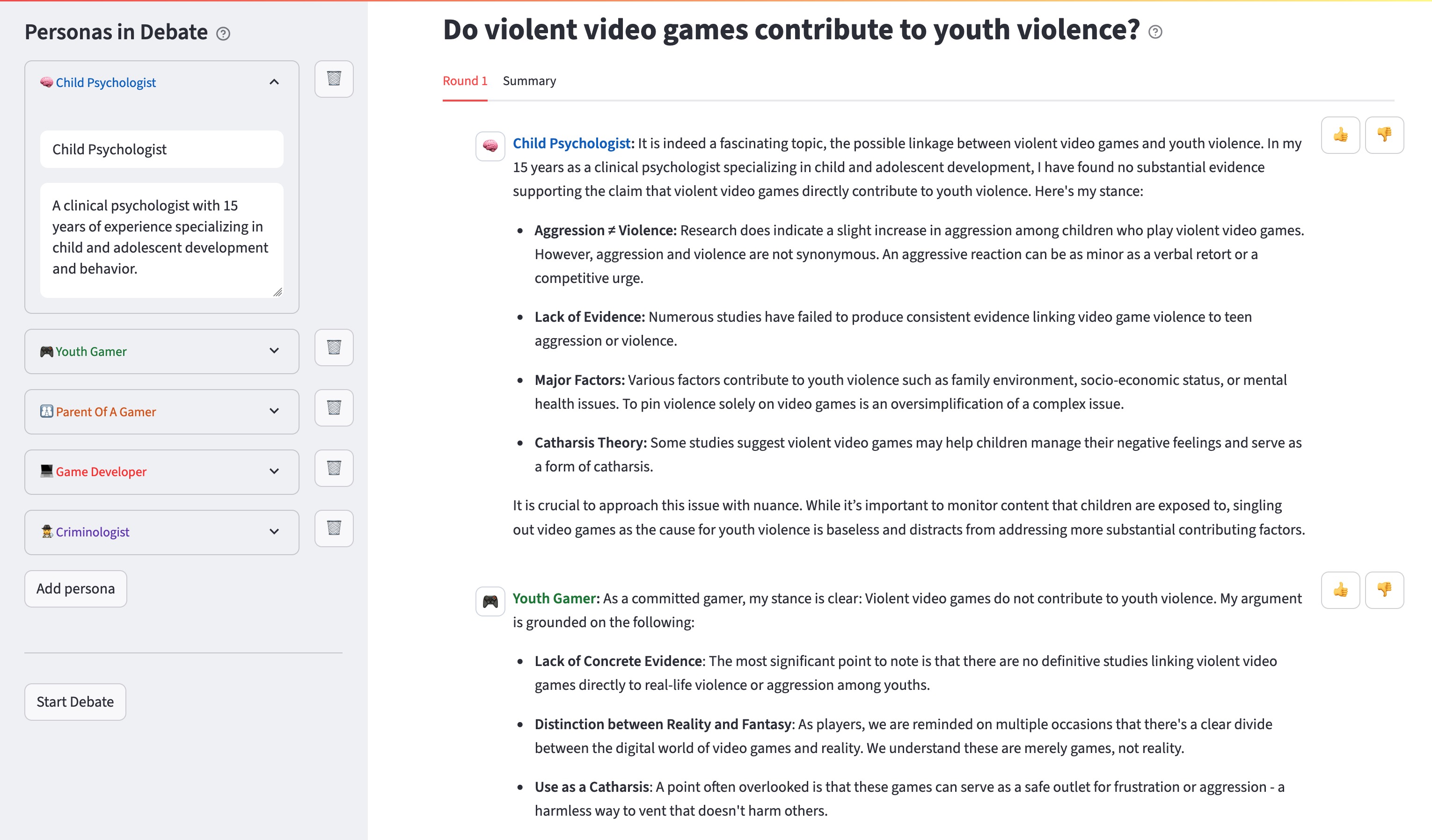}
    \caption{A screenshot of the experimental multi-persona debate system. LLM personas are shown on the left, which users can freely modify, replace, or delete. Persona arguments appear on the right, organized in a debate thread. Once the debate starts, each argument has thumbs up/down buttons to the top-right corner. Users can use them to bookmark arguments for later review and indicate whether the argument resonated with their own thinking. Bookmarked arguments are collected for the user's review in the Summary tab.}
    \label{fig:experiment}
\end{figure}

\subsubsection{Experimental LLM-generated multi-persona debate system}
For the experimental condition, we developed a multi-persona, LLM-based debate system (Figure \ref{fig:experiment}). Similar to previous design work, the system supports exploring topics through structured and dynamic dialogue among AI-generated personas \citep{Park2025-mx} and allows persona customization \citep{Salminen2024-ik}. Below, we briefly describe the user experience and technical implementation of this system.

\textbf{User experience:} Users begin by entering or selecting a topic, after which the system generates a predefined set of diverse personas to represent key perspectives on the issue. 

Each persona is presented with a title, description, profession, and visual identifiers such as colors and icons. Users can further customize, add, or edit personas to ensure comprehensive viewpoint coverage before the debate begins. Debates proceed in a round-table style, with each persona offering one argument per round in a sequential, dialogue-like manner. In the first debate round, every persona delivers an opening statement clarifying their stance and putting forward supporting arguments, structured as bullet points. Once all personas have shared their opening statements, they take turns presenting counterarguments in subsequent rounds. Users can initiate new rounds, change personas between rounds, and navigate across rounds using tab-based navigation.

To reduce cognitive overload and support reflection, the system allows users to bookmark impactful arguments and record their reactions using up/down thumbs. Bookmarked arguments are aggregated in a summary tab for easy access throughout the interaction.

The content of debate arguments varies naturally depending on the personas selected. Expert professional personas tend to produce formal arguments, while everyday stakeholder personas generate more conversational responses. This variability represents an inherent characteristic of persona-driven systems and reflects how such systems would realistically operate. More broadly, the debate system and the retrieval baseline are two distinct types of information system that people would actually use, differing in characteristics, such as persona-based, conversational, multi-round interaction on one side and source-grounded, statically presented evidence on the other. From an ecological validity standpoint, we therefore compare the two as complete systems, as users would encounter them in practice, and interpret our findings at the level of the systems as a whole.

\textbf{Technical implementation.} We implement the system using prompt engineering with LLM back-end API services and an interactive front-end to support dynamic, diverse persona-driven debates. 

We ran the system using OpenAI's GPT-4\footnote{GPT-4: \url{https://openai.com/index/gpt-4/} 
was one of the most capable LLMs at the time of our study.} and built it using Streamlit\footnote{\url{https://streamlit.io/}}, enabling fast, interactive prototyping and user-driven interface refinement. While we used GPT-4 in our implementation, our architecture is agnostic to the particular LLM used. Prior to deployment, we conducted an internal assessment comparing our GPT-4 pipeline against two alternative configurations: GPT-3.5 via the OpenAI Assistants API, and Cohere's RAG pipeline supported by their Command model and in-line citation support. Across factual accuracy and argumentation quality, our GPT-4 pipeline produced comparable outputs, supporting its selection as the primary backend (see~\ref{appendix:llm-assessment} for details).

As documented in the ~\ref{appendix:llm-assessment}, we also integrated a retrieval-augmented generation (RAG) component into the system to ground persona responses with external evidence. However, in our pilot study, we found that RAG increased response generation time, which prolonged user dwell time and disrupted the natural distribution of visual attention captured by our eye-tracking measures. To ensure smoother reading experiences and obtain more consistent attention data, we decided to abandon RAG for the main study. Nonetheless, RAG represents a more ideal approach for mitigating potential LLM hallucinations.

In the following sections, we refer to our multi-persona debate system as the \textbf{debate system} for brevity.

\subsection{Measurements} \label{subsec:measurement}


\subsubsection{Visual Attention} Drawing on confirmation bias from the cognitive psychology literature (Section \ref{subsec:biases}), confirmatory visual attention (i.e., specifically, heightened attention to content aligning with user prior beliefs) serves as a behavioral manifestation of confirmation bias in selective exposure \citep{betsch2001effects, jonas2008path}. We operationalize visual attention as participants' eye fixations (fixation duration and count). In our study, confirmatory visual attention refers to participants' eye fixation on attitude-consistent content, representing a behavioral manifestation of confirmation bias.

To quantify this, we used a two-step approach. First, we combined eye-tracking data, user pre-task beliefs, and the stance of each content item to identify whether user visual attention was directed toward information consistent or inconsistent with prior beliefs. Second, we calculated the visual attention ratio to assess the degree of focus on attitude-consistent versus -inconsistent content.

\textbf{Step 1: Annotating Gaze Patterns.} We calculated \textbf{Total Fixation Duration} and \textbf{Total Fixation Counts} from the eye-tracking records. Fixations were filtered to include only high-quality fixations, lasting between 100 and 2000ms. \textbf{User belief} was collected in the pre- and post-task questions and measured via a 5-point Likert scale: very supportive, somewhat supportive, neutral, somewhat opposed, and very opposed. To identify participant's attention to attitude-consistent versus attitude-inconsistent content, we labeled \textbf{Areas of Interests (AOIs)} based on whether the stance of the content matches user beliefs (see Table \ref{tab:belief_consist}). For example, if the content supports the given topic and the participant also holds a supportive view, then the content would be labeled as attitude-consistent information.

\begin{figure}[ht]
    \centering
      \begin{subfigure}[b]{0.47\textwidth}
        \centering
        \includegraphics[width=\textwidth]{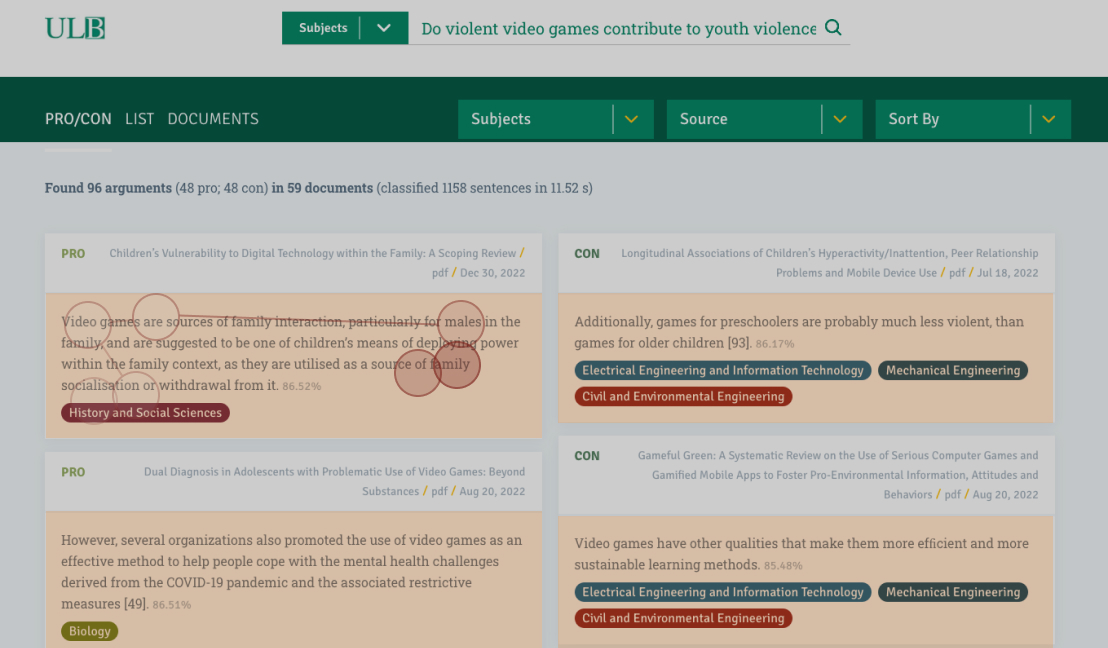}
        \label{fig:image1}
      \end{subfigure}
      \begin{subfigure}[b]{0.47\textwidth}
        \centering
        \includegraphics[width=\textwidth]{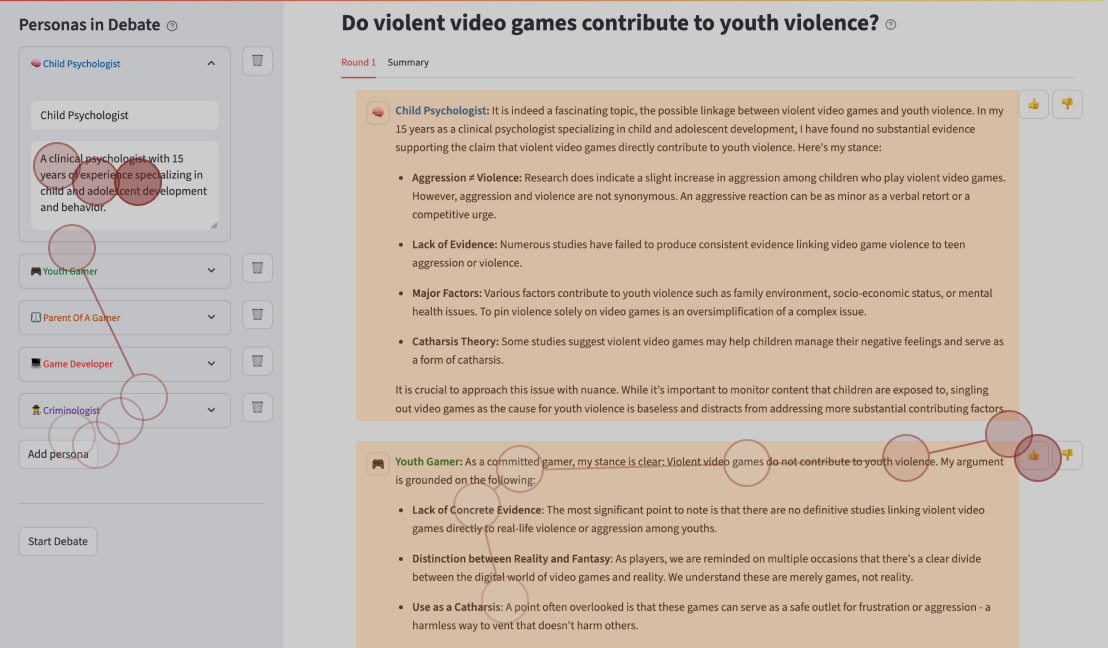}
        \label{fig:image2}
      \end{subfigure}
      \caption{An example of annotated AOI. AOI size was measured in pixels, calculated as the total number of pixels within each designated area of interest on screen. For the baseline system, AOI size corresponded to the screen region occupied by each pro or con content snippet. For the debate system, AOI size corresponded to the screen region occupied by each persona argument block.}
    \label{fig:valid visual attention}
\end{figure}

In our baseline system, attitude-consistent and attitude-inconsistent content were displayed side-by-side, allowing us to automatically assign AOIs for eye-tracking to each content type. In contrast, the debate system’s round-table format required manual AOI annotation for the stances of the argument. For each conversation thread, we labeled arguments as either attitude-consistent or -inconsistent (see Figure~\ref{fig:valid visual attention}). Arguments generated by personas with a neutral stance were excluded from AOI annotation, as a neutral persona argument cannot be meaningfully classified as attitude-consistent or attitude-inconsistent relative to any participant's prior belief. Similarly, trials in which a participant reported a neutral pre-task stance (i.e., a score of 3 on the 5-point belief scale) were excluded from the non-confirmatory visual attention ratio calculation, as the attitude-consistent vs. attitude-inconsistent distinction becomes undefined when no clear prior belief exists \citep{Nickerson1998-sl}.
Inter-coder reliability for AOI stance classification was assessed on a randomly selected 10\% of AOIs, independently coded by a second rater. Agreement was almost perfect (Cohen's $\kappa = .91$, percent agreement $= .94$), indicating high reliability of the annotation scheme. Disagreements were resolved through discussion between the two coders.

We decided not to control for minor UI differences between the two interfaces (e.g., color schemes and argument layout) in our eye-movement analysis. This decision is grounded in established evidence showing that eye movements are driven primarily by high-level cognitive factors rather than low-level visual features. Classic work by \citet{Yarbus1967-xf} demonstrated that gaze patterns during visual tasks are strongly influenced by viewers’ goals and task instructions, rather than by purely sensory properties of the image. Later studies, such as the review by \citet{Tatler2010-vz}, reinforced the conclusion that top-down task demands systematically guide attention. Empirical findings by \citet{Einhauser2008-dr} further show that, in complex visual environments, cognitive goals outweigh bottom-up saliency cues in determining fixation behavior. Given that our participants were engaged in a clearly defined decision-making task, we therefore expect their eye movements to be primarily guided by task goals and argumentative content, rather than by superficial visual differences across interfaces.



\begin{table}[ht]
\centering
\begin{tabular}{@{}llcc}
\toprule
\multicolumn{2}{l}{ } & \multicolumn{2}{c}{Pre-task Belief of Topics} \\ \cline{3-4}
\multicolumn{2}{l}{} & \begin{tabular}[c]{@{}c@{}}Very Supportive /\\Somewhat Supportive\end{tabular} & \begin{tabular}[c]{@{}c@{}}Very Opposed /\\Somewhat Opposed\end{tabular} \\ 
\midrule
\multirow{2}{*}{Content Stance} & PRO & consistent & inconsistent \\ \cline{2-4}
 & CON & inconsistent & consistent \\ 
\bottomrule
\end{tabular}
\vspace{0.2cm}
\caption{Belief consistency of areas of interest (AOIs) in relation to pre-task belief and content stance. In the baseline system, content stance was categorized as PRO or CON. This corresponds to the same dimension used for participant beliefs, where stances are labeled as supportive or opposed.}
\label{tab:belief_consist}
\end{table}

\textbf{Step 2: Quantifying Visual Differences.} We computed a \textbf{Non-Confirmatory Visual Attention Ratio} to quantitatively assess how participants distributed their visual attention between attitude-consistent and attitude-inconsistent content. This ratio is defined as the proportion of eye-tracking metrics (total fixation duration and, separately, counts) on attitude-inconsistent content relative to the total fixation activity for both attitude-consistent and inconsistent content (Equation~\ref{eq:visual_attention_ratio}). \textbf{If participants spend more time on reading attitude-inconsistent content, this ratio is > 0.5. Additionally, if the debate system effectively reduces visual attention towards attitude-consistent content, this ratio should be higher compared to the baseline system for both fixation duration and counts.}

\begin{equation}
\begin{adjustbox}{max width=0.9\textwidth}
$
\text{Non-Confirmatory\_Visual\_Attention\_Ratio}
=
\frac{
    \dfrac{\text{Fixation}_{\text{inconsistent}}}{N_{\text{inconsistent}}}
}{
    \dfrac{\text{Fixation}_{\text{inconsistent}}}{N_{\text{inconsistent}}}
    +
    \dfrac{\text{Fixation}_{\text{consistent}}}{N_{\text{consistent}}}
}
$
\end{adjustbox}
\label{eq:visual_attention_ratio}
\end{equation}

To ensure a fair comparison between interfaces and further reduce potential sensory-driven confounds, we normalized both total fixation duration and counts by the number of personas (\textit{N}) and the size of AOIs when calculating the ratio. 
Normalization was performed in two steps. First, fixation metrics (total fixation duration and fixation count) within each AOI were divided by the AOI size in pixels, yielding a fixation density measure that accounts for differences in the physical size of content areas across the two interfaces. This step was necessary because the interfaces differed substantially in their UI layout and screen design, including content width, text density, and the spatial arrangement of information. As a result, raw fixation totals might otherwise reflect differences in the amount of screen space occupied by content rather than genuine differences in attentional allocation.
 
Second, for the debate system only, the normalized fixation metrics were further divided by the number of personas ($N$) presenting pro and con arguments, respectively, thereby equalizing the contribution of each side regardless of how many personas represented it. This adjustment accounts for the fact that the debate system presents multiple content snippets distributed across several personas per round, whereas the baseline system presents a fixed and equal number of pro and con snippets. Without this normalization, conditions with more personas would artificially inflate fixation totals.

Together, these two normalization steps ensure that the Non-Confirmatory Visual Attention Ratio reflects genuine differences in attentional allocation rather than structural differences between the two interfaces.


\subsubsection{Belief Change}
Assessing attitudinal change through pre- and post-test self-reports is a well-established approach for examining how interventions influence user information-seeking behavior. This method has been used both in studies of a single, immediate system interaction \citep{Rieger2024-iq} and in longitudinal survey designs \citep{Kowalski2023-ku}. It has also been applied to measure belief change in comparable multi-persona debate settings \citep{Song2025-of}.

Building on the standard pre- and post approach, we made two theoretically motivated adjustments in computing the \textbf{Belief Change Score}. Specifically, we weighted stance change by confidence and excluded cases originating from a neutral stance (Equation~\ref{eq:belief_change}). 

First, we weighted each stance change by the participant's confidence. Controversial beliefs are often highly resistant to change. A shift away from a weakly held stance is comparatively easy and provides limited evidence of debiasing, whereas a shift away from a firmly held stance represents precisely the resistant case that is diagnostic of reduced confirmation bias. We therefore computed the confidence weight by averaging each participant's reported confidence in their stance before and after the intervention, measured on a 5-point Likert scale (i.e., very low, slightly low, moderate, slightly high, very high).

Second, we excluded cases in which the pre-intervention stance was neutral, as the metric is intended to capture shifts away from an existing stance rather than movement from a neutral position. This choice follows from the distinction between opinion formation and belief polarization. Prior work by \citet{Song2025-of} measures belief change by counting cases in which participants move from a neutral position to any clear stance afterward, but this method is not appropriate for studying confirmation bias or belief polarization. In polarization research, the critical question is whether people become more extreme in a direction they already leaned, not simply whether they move away from neutral. Including originally neutral cases would artificially inflate belief change scores and could wrongly suggest the presence or absence of confirmation bias, when participants may have simply formed an informed opinion or reflected a positivity congruency rather than become more extreme \citep[pp.~200--201]{Nickerson1998-sl}. This restriction also helps ensure that even small observed changes reflect movement from a pre-existing stance and are more meaningful for assessing bias mitigation.

Together, these two choices provide a more reliable indicator of bias mitigation than treating all stance changes equally. \textbf{Based on the compute, a positive score indicates that the system led participants to shift their stance away from their original position (e.g., from supportive to opposing), whereas a negative score reflects reinforcement of the same side of the stance.} Notably, this belief change represents a short-term attitudinal outcomes associated with the experimental interaction rather than evidence of lasting attitude conversion.

\begin{equation}
\begin{adjustbox}{max width=0.9\textwidth}
$
\text{Belief\_Change} =
\begin{cases}
(\text{pre\_stance} - \text{post\_stance}) \times \dfrac{\text{pre\_confidence} + \text{post\_confidence}}{2}, & \text{if } \text{pre\_stance} > \text{neutral} \\
(\text{post\_stance} - \text{pre\_stance}) \times \dfrac{\text{pre\_confidence} + \text{post\_confidence}}{2}, & \text{if } \text{pre\_stance} < \text{neutral} \\
0, & \text{if } \text{pre\_stance} = \text{neutral}
\end{cases}
$
\end{adjustbox}
\label{eq:belief_change}
\end{equation}

\subsubsection{Cognitive Tendencies Towards Bias and Topic Familiarity} Our study also incorporates two individual factors contributing to confirmation bias: individual cognitive tendencies towards bias and topic familiarity. To measure the first one, which indicates the individual degree of open-mindedness, we used the questionnaire developed by \citet{berthet2021measurement, Berthet2022-mx}. Participant scores were divided into three levels (low, medium, and high), where higher levels indicated less open-mindedness and a stronger inclination toward confirmatory information. To measure topic familiarity, we used a 5-point Likert scale ranging from unfamiliar to very familiar (i.e., unfamiliar, slightly familiar, moderately familiar, familiar, very familiar).




\subsection{Research Questions and Hypotheses} \label{subsec:hypothesis}

\subsubsection{Study Goal}
Our study aims to evaluate whether the debate system provides substantive benefits over established debiasing strategies, as reflected in its effectiveness in reducing confirmatory visual attention and shifting strongly held prior beliefs. Although empirical evidence regarding the effectiveness of multi-persona debate systems remains limited, recent findings suggest their potential to mitigate confirmation bias (Section~\ref{subsec:llm simulation}). We treat this optimism as a claim to be tested rather than assumed. 

We operationalize this test primarily through visual attention allocation, measured via eye-tracking (Section~\ref{subsec:measurement}). This measure provides a direct comparison between the two systems so the behavioral difference lies in how attention is distributed across the opposing content rather than in which content is encountered. By pairing this behavioral measure with belief change, we can examine whether increased attention to non-confirmatory content is associated with subsequent changes in participant beliefs. This allows us to assess confirmation bias mitigation through its downstream influence on beliefs, rather than assuming that greater visual attention alone constitutes evidence of mitigation.

We formulated hypotheses to examine whether the debate system outperforms the opposing-stances retrieval system and whether it attenuates the influence of individual factors associated with confirmation bias. Framing the hypotheses allows us to evaluate the optimistic claim from prior work directly. If hypotheses are not supported, the null results themselves constitute informative evidence about under which conditions multi-persona debate may or may not provide benefits.

Specifically, we hypothesize that: \textbf{compared to the opposing-stances retrieval-based system, the multi-persona debate system will more effectively mitigate confirmatory visual attention and promote belief change}. We further hypothesize that \textbf{individual factors associated with confirmatory attention, such as topic familiarity and cognitive tendencies, will exert weaker effects in the multi-persona debate condition}. 

\subsubsection{Research Model}
\begin{figure*}[ht]
    \centering
    \includegraphics[width=0.6\textwidth]{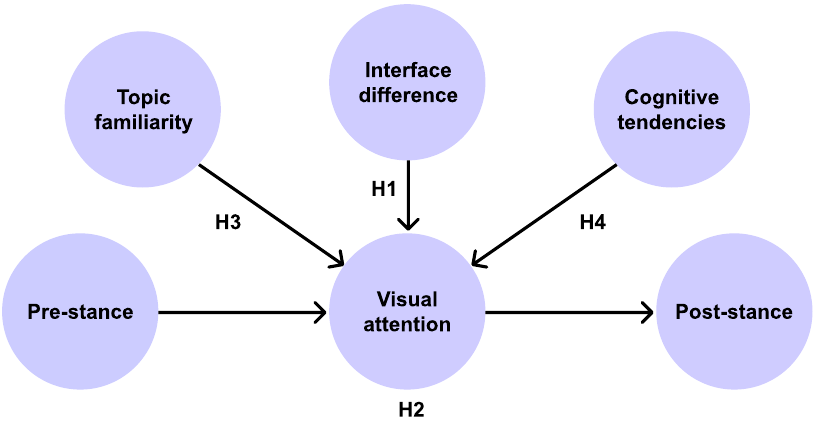}
    \caption{Research model illustrating the hypothesized relationships among system conditions, visual attention, and belief change, moderated by individual factors, including topic familiarity and cognitive tendencies.}
    \label{fig:research_model}
\end{figure*}

To clarify our hypotheses, we present our research model in Figure~\ref{fig:research_model} and formulate detailed hypotheses using variables defined in the measurements (Section~\ref{subsec:measurement}). We organized these hypotheses under two primary research questions.

\begin{description}
    \item[RQ1] Compared to the opposing-stances retrieval system, can the debate system better reduce confirmatory visual attention and promote belief change?
    \begin{description}
        \item[H1] User non-confirmatory visual attention ratio (i.e., the proportion of visual attention towards attitude-inconsistent content) is expected to be higher in the debate system than in the baseline system.
        \item[H2] A higher non-confirmatory visual attention ratio will lead to a higher belief change score in the debate system compared to the baseline system.
    \end{description}
    \item[RQ2] Compared to the opposing-stances retrieval system, can the multi-persona debate system better counteract visual bias influenced by individuals' topic familiarity and cognitive tendencies towards bias?
        \begin{description}
        \item[H3] Users with higher topic familiarity will exhibit a lower non-confirmatory visual attention ratio (i.e., visual attention leans towards attitude-consistent content). However, the debate system could counter this tendency.
        \item[H4] Users with stronger cognitive tendencies towards bias will exhibit a lower non-confirmatory visual attention ratio. However, the debate system could counter this tendency.
    \end{description}
\end{description}

\subsection{Dataset}

\subsubsection{Overview}
As shown in Table~\ref{tab:dataset_overview}, we included data from 39 of the 40 participants after excluding one participant whose eye-tracking data failed the quality check. The quantitative data comprised 312 system trials with eye-tracking measures of visual attention, belief change scores from pre- and post-task surveys, trial-level topic familiarity ratings from the pre-task survey, and participant-level measures of cognitive tendencies toward bias collected before the task.

\begin{table}[h!]
\small
\centering
\caption{Quantitative Dataset Overview}
\label{tab:dataset_overview}

\begin{tabular}{@{}>{\RaggedRight}p{2cm} >{\RaggedRight}p{3.5cm} >{\RaggedRight\arraybackslash}p{8cm}@{}}
\toprule
\multicolumn{3}{c}{\textbf{Category}} \\
\midrule

\multirow{4}{*}{\textbf{Sample}} 
& Participants & 39 valid \\
& Design & Within-subject \\
& Trials per participant & 8 controversial topics \\
& Total trials & 312 (39 $\times$ 8) \\

\midrule
\multirow{2}{*}{\textbf{Conditions}}
& Baseline &  Opposing-stances retrieval \\
& Experimental & Multi-persona debate \\

\midrule
\multirow{4}{*}{\textbf{Measurable}} 
& Eye-tracking record &  Visual attention metrics (per trial) \\
& Belief changes & Pre- and post-intervention belief (per trial) \\
& \multirow{2}{*}{Individual differences} & Topic familiarity (per trial) \\
& & Cognitive tendencies toward bias (per participant) \\
\bottomrule
\end{tabular}

\end{table}

\subsubsection{Power Analysis}
To determine the required sample size for hypothesis testing, we conducted a prior power analysis using G*Power \citep{Faul2007-ej} based on our within-subjects design and hypotheses. Drawing on prior work suggesting meaningful benefits of multi-persona debate systems for reducing confirmation bias, we set the expected effect size at medium ($d = 0.6$). Using a two-sided paired $t$-test with a power of 0.95 and $\alpha = .05$, the analysis indicated a minimum required sample size of \textit{39} participants. Our recruited sample of 40 participants yielded 39 valid participants after the initial data-quality check, meeting the target sample size. Subsequent hypothesis-specific exclusions reduced the participant-level sample for H1 and H2 to 37 participants.
This sample size aligns with common practice in eye-tracking research. A systematic review of 161 eye-tracking studies conducted in educational settings with similar student populations found an average sample size of about 28 participants \citep{Strohmaier2020-iv}. Our sample is also larger than what is usually seen in HCI studies. \citet{Caine2016-pz} reports that the mean sample size for usability studies in CHI2014 submissions is 17.

The prior power analysis also provides a guideline for interpreting our hypothesis results. Detecting significant differences with effect sizes above the medium threshold would inform that the debate system offers a real advantage over the baseline in attention allocation and belief change. The observed null results are themselves informative, indicating that claims about the benefits of the debate system should be treated with caution and serving as a corrective to overly optimistic expectations. This power analysis pertains specifically to the participant-level paired comparisons (H1). The hypotheses tested through regression models, the interaction between attention and interface condition on belief change (H2) and the moderating roles of topic familiarity (H3) and cognitive tendencies (H4), were not its target. We accordingly interpret those results more tentatively, particularly where the evidence is marginal.



\subsection{Mixed-Methods Analysis} \label{subsec:analysis}

We conducted a mixed-methods analysis combining quantitative and qualitative approaches to address the two research questions. For quantitative analysis, we tested hypotheses using eye-tracking and survey data by examining relationships between measured dependent and independent variables. Eye-tracking data capture visual attention, operationalized as the non-confirmatory visual attention ratio based on fixation counts and durations (see Equation \ref{eq:visual_attention_ratio}). Survey data include attitudinal stance, confidence ratings, topic familiarity (pre-task, per-trial measure), cognitive tendencies toward bias (post-study questionnaire), and belief change.

We assessed data normality using Shapiro–Wilk tests and applied parametric or non-parametric tests accordingly. For qualitative analysis, we conducted collaborative thematic analysis~\citep{Richards2018-jj} of semi-structured interviews, using open coding followed by axial coding to refine and synthesize emerging themes. The initial coding scheme is guided by the interview protocol (\ref{appendix:measurement}). We provide further analysis details for addressing each research question below.

\subsubsection{RQ1: Can the debate system better reduce confirmatory visual attention and promote belief change?}

To address RQ1, our analysis examines two distinct constructs. Eye-tracking serves as the primary behavioral measure of attention allocation (i.e., how participants distribute attention across content once encountered) operationalized as the non-confirmatory visual attention ratio. Survey data capture belief change (i.e., whether attitudes subsequently change) operationalized as the belief change score. H1 tests whether the debate system leads to higher non-confirmatory visual attention compared to the baseline using a two-sided paired $t$-test at the participant level. H2 examines the interaction between interface condition and non-confirmatory visual attention in predicting belief change using a linear mixed-effects model. 

Whereas attention allocation and belief change can be measured consistently across both systems and therefore serve as our primary cross-system dependent variables, the debate system provides an additional opportunity to examine selective exposure as a distinct construct. In the debate interface, users actively choose which personas to interact with, making content selection an observable, user-driven behavior. In contrast, the retrieval baseline presents a fixed and balanced mix of attitude-consistent and inconsistent content, leaving no comparable selection behavior to measure. We therefore conduct an additional analysis specific to the debate condition using the non-confirmatory persona selection ratio. Analyzing this behavior separately allows us to examine whether confirmation bias emerges at the stage of content selection, prior to and conceptually distinct from how attention is subsequently distributed across the content users encounter.

We then analyzed qualitative interview data on participants' information-seeking strategies and perceptions of argument quality across the two systems. Both quantitative and qualitative results are reported in Section~\ref{subsec:quantitative_RQ1}.

\subsubsection{RQ2: Can the debate system counteract visual bias influenced by topic familiarity and cognitive tendencies?}

To address RQ2, survey data provide the moderator individual variables (topic familiarity and cognitive tendencies toward bias), while eye-tracking data again provides the measure of attention allocation. We conducted analysis at the trial level to capture within-subject variation across topics. We tested H3 and H4 using beta mixed-effects regression models with interaction terms between interface condition and each moderator. Similarly, we conducted thematic analysis from the interview data to contextualize how topic familiarity and cognitive tendencies shaped participants' engagement strategies and expectations across the two systems. Both quantitative and qualitative results are reported in Section~\ref{subsec:quantitative_RQ2}.

\section{Reliability Checks} \label{sec:reliability}


Our within-subjects design introduces two potential sources of confounding that could threaten the validity of hypothesis testing: order effects and anchoring effects. Despite our use of randomization and counterbalancing to minimize these risks, some residual effects may still persist. Following the recommended practices of \citet{Cockburn2019-aw}, we conducted reliability checks to empirically verify their absence before proceeding to hypothesis testing. Additionally, we conducted and reported reliability and psychometric analysis of the survey results.

\subsection{Potential Confounds and Mitigation Strategies}

\textbf{System order effects} refer to whether the sequence in which participants experienced the two interface conditions influenced their responses, such as if exposure to the retrieval system first systematically inflated or deflated subsequent responses to the debate system. We address this by using random assignment and a counterbalanced parallel control condition within a within-subjects design. Participants were also given a 15-minute interval to transition to using the other system. We performed an asymmetric transfer check (Section~\ref{sec:asymmetric}).

\textbf{Topic order effects} refer to whether the position of a topic within a block influenced responses independently of topic content. These are mitigated by design through within-block topic randomization, and further accounted for statistically in Section~\ref{subsec:quantitative_RQ2}, where topic is included as a random effect $(1 | topic_id)$ in hypothesis testing, capturing residual variance attributable to individual topics.

\textbf{Anchoring effects} refer to whether participants' responses became disproportionately anchored to their initial exposures when individual factors, specifically topic familiarity and cognitive tendencies towards bias, interact with system order. We performed an anchoring check (Section~\ref{sec:anchoring}).


\subsection{Detecting Asymmetric Transfer} \label{sec:asymmetric}
We calculated stance difference (i.e., post\_stance minus pre\_stance) to capture any changes in beliefs across all trials. This score is suitable for reliability tests by incorporating all trials compared to the belief change score (Equation \ref{eq:belief_change}). In our case, asymmetric transfer would be indicated by a statistically significant stance difference between the two group orderings in which the mean stance difference for the Block A/B group significantly differs from that of the Block B/A group.

As the stance difference did not meet the assumption of normality, we used aligned rank transform (ART) ANOVA. The model followed the formula: $\text{Stance\_difference} \sim \text{order} \times \text{interface} + (1 \mid \text{user\_id}).$ 

\textbf{Result: No asymmetric transfer detected.}
The result only revealed a significant influence of interface conditions, \( F(1, 278) = 10.24, p = .0015 \), indicating that pre–post stance differences existed depending on the interface used. The order influence was not significant, \( F(1, 38) = 0.33, p = .57 \), nor was the interaction between the order and interface conditions, \( F(1, 278) = 1.56, p = .21 \). These results suggest no evidence of an asymmetric transfer effect.

\subsection{Detecting Anchoring on Independent Factors} \label{sec:anchoring}
Two primary independent variables (i.e., topic familiarity and cognitive tendencies towards bias) may contribute to anchoring effects. We examine whether stance difference is disproportionately influenced by both factors in the presence of order exposure to different interfaces. If order effects are found to be significant, it would suggest that any observed mitigation effects are due not to the interface itself, but rather to the sequence in which participants experienced the different interfaces.

Because the stance difference data did not meet the normality assumption, we used a four-way ART ANOVA as a non-parametric test with topic familiarity, cognitive tendencies, order, and interface as factors.

\textbf{Result: No anchoring effects detected.}
The result revealed a significant main effect of interface on stance difference, \(F(1, \allowbreak 251.39) = 7.96, p = .0050 \). No main effects were found for order, familiarity, or cognitive tendencies. Additionally, the interaction between order and interface was also not significant, \( F(1, 260.58) = 0.0008, p = .9780 \). However, the interaction effect between topic familiarity and cognitive tendencies towards bias was significant \( F(8, 259.54) = 2.87, p = .0045 \). This partially indicates the joint influence of topic familiarity and cognitive tendencies on belief change. Since the dependent variable in this reliability analysis is stance difference rather than the actual belief change score (i.e., equation \ref{eq:belief_change} that excludes trials with prior belief in neutral stance), we present further analysis in Section~\ref{sec:results} to fully test our hypothesis.

Finally, a correlation analysis between topic familiarity and cognitive tendencies yielded a weak but statistically significant correlation \( (r = 0.18, p < .05)\). While the relationship is statistically significant, the magnitude of the association is weak, indicating that the two factors are largely independent and unlikely to confound each other.

\subsection{Reliability and Psychometric Analysis on Survey Results}
We conducted standard reliability and descriptive analysis to evaluate the psychometric properties of the survey results. Cognitive tendency was measured using a four-item forced-choice scale (Cronbach’s $\alpha = .80$, mean inter-item $r = .50$), with corrected item–total correlations ranging from .57 to .73, indicating good internal consistency. Evidence for construct validity comes from the original scale development by \citet{berthet2021measurement}, who demonstrated that the measure reliably captures stable individual differences in confirmatory thinking dispositions. Cognitive tendency scores showed an observed range of 0.28–0.75 ($M = 0.52$, $SD = 0.14$, skewness $= 0.07$, kurtosis $= -1.20$), suggesting an approximately normal distribution without evidence of floor or ceiling effects. Topic familiarity (coded 1 = \textit{not familiar} to 5 = \textit{very familiar}) had a mean of $M = 2.75$ ($SD = 1.17$, skewness $= 0.12$, kurtosis $= -0.87$). These values show a well-distributed range of responses with no apparent floor or ceiling effects. We provide the full questionnaire in ~\ref{app:cb_survey}.

\section{Findings} \label{sec:results}

\subsection{Compared to the Opposing-stances Retrieval system, Can the Multi-Persona Debate System Better Reduce Confirmatory Visual Attention and Promote Belief Change?} \label{subsec:quantitative_RQ1}

To test \textit{H1} and \textit{H2}, we excluded trials where participants did not have a clear prior stance (i.e., neutral) on the topic. We also removed trials with visual attention ratios of 0 or 1. These extreme values indicate that participants only viewed arguments from one side of the debate, reflecting selection bias rather than strictly attentional bias. Including these trials would have biased comparisons with the baseline system, where arguments from both sides were always presented. After incorporating data from participants who selected personas exclusively from one side, we found a slight participant preference for selecting attitude-consistent personas. After these exclusions, 225 valid trial-level observations remained. For formal hypothesis testing, we averaged the results at the participant level, yielding 37 data points.

In the test for normality, all variables passed the Shapiro-Wilk test: non-confirmatory visual attention ratio based on fixation counts (\(W = 0.9664, p = .3194\)) and fixation duration (\(W = 0.9489, p = .0893\)), and belief change score (\(W = 0.9696, p = .3993\)). Accordingly, to test \textit{H1}, we conducted a two-sided paired \textit{t}-test to compare the mean differences in non-confirmatory visual attention ratios between the two systems. To test \textit{H2}, we employed a linear mixed-effects model to examine whether the attention ratio influenced the effect of system condition on belief change: \text{Belief\_change} $\sim$ \text{Non-confirmatory\_attention\_ratio} (\text{duration} / \text{counts}) $\times$ \text{interface} + (1 $\mid$ \text{user\_id}).




\subsubsection{Both Systems Showed Greater Visual Attention to Attitude-Inconsistent Content, with No Significant Difference Between Them} \label{quant:oppose H1} 

\begin{figure}[ht]
    \centering
    \includegraphics[width=0.8\textwidth]{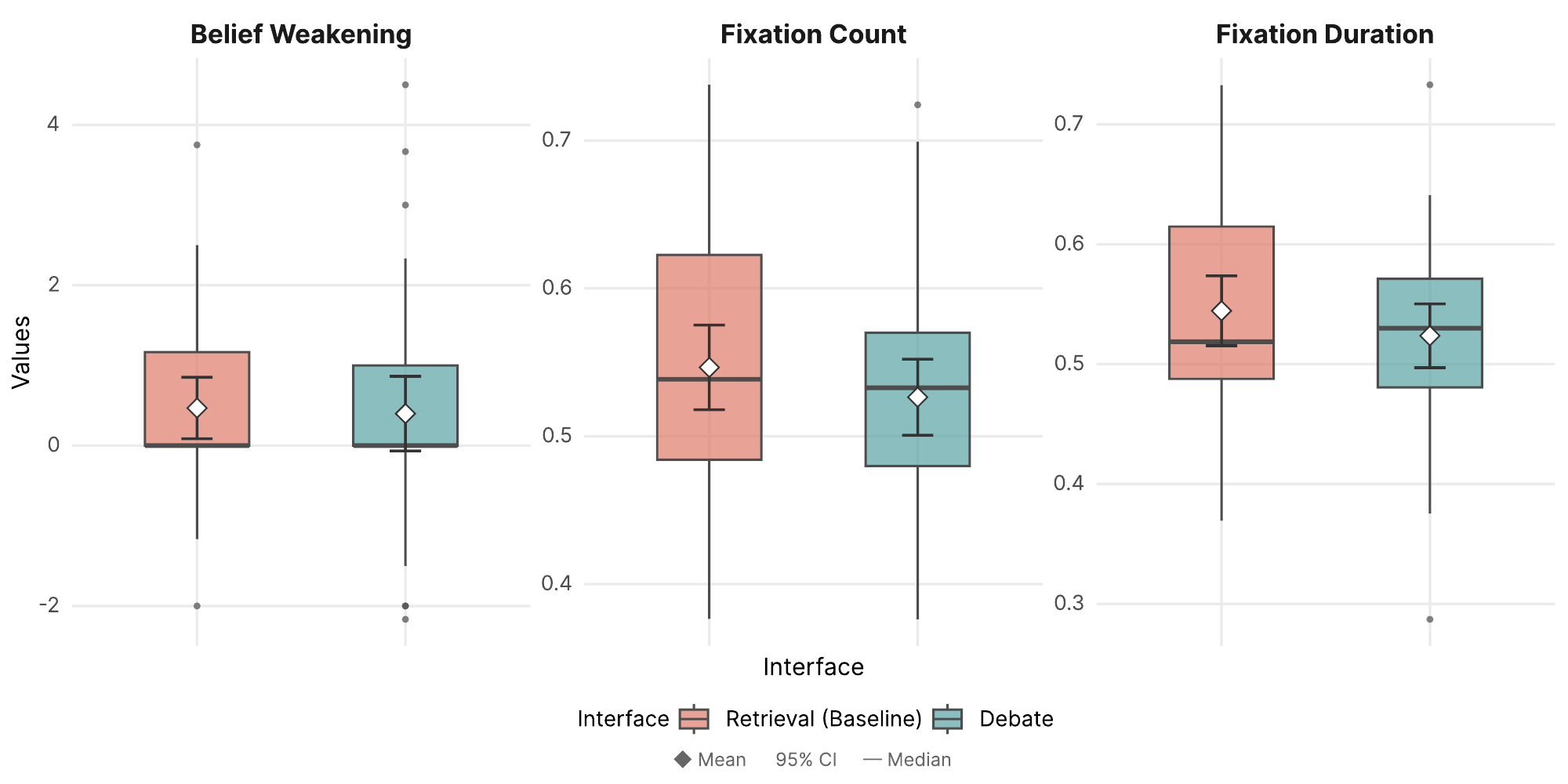}
    \caption{Boxplots of belief change score, non-confirmatory fixation count ratio and duration ratio across interface conditions. Boxes show the interquartile range; horizontal lines indicate medians; diamonds indicate means with 95\% confidence intervals; dots represent outliers.}
    \label{fig:boxplot_variables}
\end{figure}

\textbf{Hypothesis Testing.} As shown in Fig \ref{fig:boxplot_variables} (the 2nd and 3rd subplots) and Table \ref{tab:summary_stats}, the mean value of non-confirmatory visual attention ratio, i.e., counts (\(M_{baseline} = 0.5466, M_{debate} = 0.5264\)) and duration (\(M_{baseline} = 0.5445, M_{debate} = 0.5237\)), were both above 0.5 across systems. This suggests that \textbf{within both experimental conditions, participants allocated more than half of their visual attention to attitude-inconsistent content}. 

\begin{table}[ht]
\centering
\resizebox{\textwidth}{!}{%
\begin{tabular}{@{}lccccccr@{}}
\toprule
\textbf{Variables} & \textbf{Baseline (Mdn / M)} & \textbf{Debate (Mdn / M)} & \textbf{$t$} & \textbf{$df$} & \textbf{$p$} & \textbf{Cohen's $d$} & \textbf{[95\% CI] for Mean} \\
\midrule
Fixation Count Ratio    & 0.5386 / 0.5466 & 0.5328 / 0.5264 & $-0.961$ & 36 & $.343$ & $-0.158$ & $[-0.063, 0.022]$ \\ 
Fixation Duration Ratio & 0.5187 / 0.5445 & 0.5300 / 0.5237 & $-0.985$ & 36 & $.331$ & $-0.162$ & $[-0.064, 0.022]$ \\
Belief Change           & 0.0000 / 0.4685 & 0.0000 / 0.3986 & $-0.288$ & 36 & $.775$ & $-0.047$ & $[-0.561, 0.422]$ \\
\bottomrule
\end{tabular}
}
\begin{flushleft}
\fontsize{8pt}{9pt}\selectfont
Two-sided paired $t$-tests with two-tailed 95\% CIs reported for the mean difference. Cohen's $d$ computed as $t / \sqrt{n}$ ($n = 37$). Effect size thresholds following \citet{Lovakov2021-dh}: small ($d = 0.15$), medium ($d = 0.36$), large ($d = 0.65$).
\end{flushleft}
\caption{Summary statistics (median, mean, paired $t$-test, and effect size) for each dependent variable across interface conditions. Non-confirmatory visual attention ratios exceeded 0.5 for both interfaces, with no significant difference. Belief change scores were positive for both conditions, with no significant difference. Results indicate participants allocated more attention to attitude-inconsistent content and shifted away from initial beliefs regardless of interface type, with neither interface outperforming the other.}
\label{tab:summary_stats}
\end{table}

In H1, we hypothesized that the debate system would be more effective in increasing users' non-confirmatory visual attention ratio than the baseline. However, the paired \textit{t}-test results showed no statistically significant difference. For fixation counts, the difference was not significant (\(t = -0.9607\), \(p = .3430\)), and for fixation durations, the result was also non-significant (\(t = -0.9851\), \(p = .3311\)). Additionally, the mean difference in attention ratio was slightly negative in the debate system, with \(-0.021\) for fixation duration and \(-0.020\) for fixation counts. The corresponding effect sizes were also small ($d = -0.158$ for fixation counts and $d = -0.162$ for fixation duration). According to empirically established benchmarks, these values lie within the small-effect range ($d = 0.15$–$0.36$) of social psychology studies related to attitudes and prejudice \citep{Lovakov2021-dh}. These results suggested that \textbf{the debate system did not increase the non-confirmatory visual attention ratio compared to the baseline. Thus, H1 was not supported.} 

\textbf{Qualitative results.} Through participant interviews, we found that participants in both systems actively sought diverse viewpoints, helping explain the similar visual attention ratios. \textbf{Across both systems, participants reported deliberately seeking diverse perspectives. This suggests that the similar visual attention ratios between both systems reflect an intentional information-seeking strategy rather than a system-driven effect}. When using the debate system, some participants selected balanced personas. P6 said ``\textit{I tried to equalize the number of people who supported and opposed, so it was easier to understand both perspectives.}'' P27 shared a similar viewpoint as ``\textit{I put together personas that would likely disagree with each other to see how they responded.''} Moreover, P34 explicitly changed persona descriptions to produce content that reflected alternative viewpoints from a specific angle: ``\textit{I set up a persona as a foreigner with no connection to America and wanted to observe how the system would respond [to the topic of `should the US implement a universal basic income'].''}

In the baseline system, which presents content from both sides of a topic, participants also reported actively jumping between content snippets to seek diverse viewpoints. P36 described this as a horizontal scanning pattern ``\textit{I wasn't reading vertically but going horizontally, moving between pro and con content.}'' Other participants also echoed this effort to achieve balance in their reading. P5 said they were ``\textit{trying to cover both as much as I could,}'' and P39 reflected, ``\textit{Rather than trying to find things that confirm what I already believe, [I wanted to] try and see, is there a good argument on the other side?}''

\textbf{Persona selection analysis}.
However, it is important to note that our visual attention analysis excluded trials in which participants selected personas from only one side in the debate system. We performed additional analysis by including these excluded trials and observed a modest preference for selecting attitude-consistent persona. We present details below.

Following the same logic as the non-confirmatory visual attention ratio, we computed a \textit{non-confirmatory persona selection ratio}, defined as the proportion of attitude-inconsistent personas out of all personas selected in a given trial. As reported in Table~\ref{tab:persona_selection}, participants selected attitude-inconsistent personas at a mean rate of 0.44 across both trial and participant levels. This value was below the unbiased benchmark of 0.50 with small and moderate effect size (trial level: Wilcoxon signed-rank test, $V = 1773.50$, $p = .007$, $r = 0.25$ and participant level: t-test, $t(38) = -2.18$, $p = .035$, $d = -0.35$), indicating a small but consistent confirmatory selection bias. However, this pattern did not amount to a fully one-sided choice. 7\% of trials relied solely on attitude-consistent personas, and another 7\% relied solely on attitude-inconsistent personas. The vast majority (86\%) combined both, but including more attitude-consistent than inconsistent personas. The observed bias therefore reflects a modest overall tilt.

\begin{table}[h!]
\small
\centering
\begin{tabular}{@{}lcccccc@{}}
\toprule
\textbf{Level} & \textbf{\textit{N}} & \textbf{\textit{M} (\textit{SD})} & \textbf{Test} & \textbf{Statistic} & \textbf{\textit{p}} & \textbf{Effect Size} \\
\midrule
Trial       & 115 & 0.44 (0.26) & Wilcoxon & $V = 1773.50$ & .007 & $r = 0.25$ \\
Participant &  39 & 0.44 (0.13) & $t$-test  & $t(38) = -2.18$ & .035 & $d = -0.35$ \\
\bottomrule
\end{tabular}
\begin{flushleft}
\fontsize{8pt}{9pt}\selectfont
In total, 115 trials are used for the persona analysis, after excluding participants with neutral positions on the topics. Depending on the data distribution, both parametric and non-parametric statistical tests are employed. 
\end{flushleft}
\caption{One-Sample Tests of Non-Confirmatory Persona Selection Ratio Against Chance (0.50)}
\label{tab:persona_selection}
\end{table}

\begin{table}[h!]
\centering
\begin{tabular}{@{}lccc@{}}
\toprule
\textbf{Measurements} & \textbf{\textit{r}\textsubscript{s}} & \textbf{\textit{S}} & \textbf{\textit{p}} \\
\midrule
Visual attention (fixation count)    & .18 & 208238 & .056 \\
Visual attention (fixation duration) & .20 & 202908 & .033 \\
Belief change                        & .38 & 158075 & $<$.001 \\
\bottomrule
\end{tabular}
\begin{flushleft}
\fontsize{8pt}{9pt}\selectfont
Correlations are calculated at the trial level (\(N=115\)). Because multiple trials are nested within participants, these correlations are reported descriptively; conventional \(p\)-values do not account for within-participant dependence.
\end{flushleft}
\caption{Descriptive Spearman Rank Correlations Between Non-Confirmatory Persona Selection Ratio and Subsequent Behaviors}
\label{tab:persona_correlation}
\end{table}

Persona selection was also associated with participants' subsequent behavior (see Table \ref{tab:persona_correlation}). Selecting more attitude-inconsistent personas showed a small positive association with non-confirmatory visual attention ratio, reaching significance for fixation duration (\textit{rho} = .20, \textit{p} = .033) but not fixation count (\textit{rho} = .18, \textit{p} = .056), and a small-to-moderate association with belief change (\textit{rho} = .38, \textit{p} < .001). Because multiple trials were contributed by each participant, these correlations are reported descriptively and their conventional \(p\)-values should not be interpreted as independent-observation inferential tests. These patterns indicate that trials involving more attitude-inconsistent persona selections tended to be associated with greater subsequent attention to attitude-inconsistent content and larger belief shifts.


Together, these results suggest that confirmation bias in the debate system can operate at an earlier, selection stage, before any attention is allocated to debate content.
Although participants in both systems spent more time reading attitude-inconsistent content overall (Table~\ref{tab:joint_display}), some participants using the debate system tended to select attitude-consistent personas more frequently than attitude-inconsistent ones. 
Excluding these trials enables a cleaner comparison of visual attention across systems, but it also removes evidence of selection bias that may arise during persona choice.


\subsubsection{Both Systems Showed Small Positive Belief-Change Scores, with No Significant Difference Between Them} \label{quant:oppose H2} 

\textbf{Hypothesis Testing.} As shown in Fig \ref{fig:boxplot_variables} (the 1st subplot) and Table \ref{tab:summary_stats}, we found that the mean belief change scores were positive for both interfaces (\(M_{baseline} = 0.4685, M_{debate} = 0.3986\)), suggesting small shifts away from participant initial stances. To examine whether these shifts differed from no belief change, we conducted two-sided one-sample t-tests against zero. As shown in Table \ref{tab:belief_change_onesample}, the mean belief-change score in the baseline was significantly greater than zero (\(t = 2.394\), \(p = .022\)). However, the mean belief-change score in the debate system was not (\(t = 1.678\), \(p = .102\)). Additionally, the difference between interfaces was not statistically significant (\(t = -0.2880\), \(p = .7750\)) with a negligible effect size ($d = -0.047$). These findings suggest that no reliable difference in belief change between the two interfaces and the debate system does not show comparative advantage than the baseline to promote belief change. Combined with H1, the minimum detectable effect size for this sample at 80\% power was $d = 0.47$, implying that any practically meaningful medium-to-large differences between the systems would likely have been observed if they were present. This finding contradicts our initial expectations. Although we observed significant stance differences between interfaces (reported in the reliability test in Section \ref{sec:reliability}), after excluding cases where participants had an initial neutral position, the final belief change did not reach a significant level.

\begin{table}[ht]
\centering
\begin{tabular}{@{}lcccccr@{}}
\toprule
\textbf{Interface} & \textbf{$M$} & \textbf{$t$} & \textbf{$df$} & \textbf{$p$} & \textbf{Cohen's $d$} & \textbf{[95\% CI] for Mean} \\
\midrule
Baseline & 0.4685 & 2.394 & 36 & .022 & 0.394 & [0.072, 0.865] \\
Debate   & 0.3986 & 1.678 & 36 & .102 & 0.276 & [$-0.083$, 0.880] \\
\bottomrule
\end{tabular}
\begin{flushleft}
\fontsize{8pt}{9pt}\selectfont
Two-sided one-sample $t$-tests examined whether participant-level
mean belief-change scores differed from zero. Positive scores indicate shifts
away from participants' initial stances. Cohen's $d$ was calculated as $M/SD$
($n = 37$).
\end{flushleft}
\caption{One-sample $t$-tests of belief change against zero across interface conditions}
\label{tab:belief_change_onesample}
\end{table}

\begin{table}[ht]
\resizebox{\textwidth}{!}{%
\begin{tabular}{@{}lcccccc@{}}
\toprule
\textbf{Factors} & 
\multicolumn{2}{c}{\textbf{Slope ($\beta$)}} & 
\multicolumn{2}{c}{\textbf{$df$ / $p$}} & 
\multicolumn{2}{c}{\textbf{Semi-partial $R^2$}} \\
\cmidrule(lr){2-3} \cmidrule(lr){4-5} \cmidrule(lr){6-7}
& Duration & Counts & Duration & Counts & Duration & Counts \\
\midrule
(Intercept) & 
$-1.701$ & $-1.531$ & 
$61.44$ / $.163$ & $61.13$ / $.218$ & 
-- & -- \\
Attention Ratio & 
$+3.848$ & $+3.524$ & 
$60.25$ / $.083$ & $59.96$ / $.118$ & 
$0$ & $0$ \\
Debate & 
$+3.233$ & $+3.459$ & 
$63.36$ / $.075$ & $64.29$ / $.066$ & 
$0.036$ & $0.039$ \\
Interaction (Attention Ratio $\times$ Debate) & 
$-6.074$ & $-6.490$ & 
$63.71$ / $.071$ & $64.60$ / $.063$ & 
$0.037$ & $0.039$ \\
\bottomrule
\end{tabular}
}
\small
We report semi-partial $R^2$ as a measure of variance explained by each predictor. Following prior work \citep{Stoffel2021-sv}, we interpret these values descriptively rather than using fixed thresholds, as no universal benchmarks exist for mixed-effects models.

\caption{Linear mixed-effects model results for belief change, with non-confirmatory visual attention ratio and interface condition as predictors. The interaction between attention ratio and interface was negative and marginally significant, indicating a suggestive pattern of a weaker attention–belief change relationship in the debate condition than in the baseline condition. The lower-order interface coefficients are evaluated when the attention ratio equals zero, which lies outside the analyzed range. Therefore, they are treated as mathematical baselines and are not substantively interpreted.}
\label{tab:lme_results_belief}
\end{table}

In H2, we aimed to understand the interaction effect between attention ratio and interface conditions on belief change. We hypothesized that the non-confirmatory visual attention ratio would more strongly influence belief change in the debate system than in the baseline. In Table \ref{tab:lme_results_belief}, results from the linear model showed no statistically significant effect of attention ratio on belief change. A marginal level of influence was observed only for fixation duration ($\beta = 3.848$, $p = .083$), whereas the effect for fixation count was not significant ($\beta = 3.524$, $p = .118$). Additionally, we observed a marginally significant increase in belief change\footnote{We use the term ``marginally significant'' for p values between 0.05 and 0.10. This choice follows the systematic review by \citet{Pritschet2016-wd} of reporting practices across psychology subfields, which finds that marginal significance is frequently reported in social psychology. Because our work lies at the intersection of HCI, information seeking, and social psychology, we adopt this convention for borderline p-values. We use the term descriptively for borderline \(p\) values while treating such results as suggestive rather than confirmatory}. when switching from the baseline to the debate system, with $\beta$ of 3.233 ($p = .075$) for fixation duration and 3.459 ($p = .066$) for fixation count. These findings suggest a positive, separate influence on belief change by either reading more attitude-inconsistent content or using the debate system. However, the interaction between attention ratio and interface was negative and marginally significant ($\beta= -6.074$, $p = .071$ for duration; $\beta= -6.490$, $p = .063$ for count). Based on the social psychology convention reported by \citet{Pritschet2016-wd}, these results are termed marginally significant, but they do not meet the conventional \(p<.05\) threshold and should therefore be regarded as suggestive rather than confirmatory. In sum, this means \textbf{reading more opposing content tended to have less impact on belief change when participants were using the debate system. Therefore, we did not find support for H2.}

\textbf{Qualitative Results.} Through participant interviews, we identified different reasons that limited belief change in the baseline and debate systems. In the baseline system, participants widely acknowledged its credibility and trustworthiness, yet explicitly noted that this did not necessarily translate into opinion shifts. For example, P26 stated that ``\textit{even though I scrolled through all of it and read beyond the abstract, definitely it's trustworthy, but I don't think it's going to affect my opinion in any sort of way.}'' Several participants attributed this to the dense presentation of information, which hindered comprehension and engagement. P25 noted that ``\textit{I didn't have enough time to go through it. The way it's formatted means I'd have to read the entire document, which would take more time and probably confuse me,}'' and P33 commented that \textit{it was just a humongous document...and the quote the system pulled out felt out of context.}''

In the debate system, while participants found the persona-driven arguments more relatable and emotionally engaging, \textbf{sustained engagement was undermined by repetitive content across rounds}. P35 noted that ``\textit{a lot of debates were reiterated...and it wasn't like new information,}'' and P26 similarly observed that ``\textit{the first round already brings up many good perspectives, but the later rounds do not.}'' Together, these patterns suggest that neither system's presentation format was sufficiently persuasive to drive meaningful belief change. The retrieval system failed due to comprehension barriers, and the debate system also failed due to diminishing argumentative novelty.

\subsection{Compared to the Opposing-stances Retrieval System, Can the Multi-Persona Debate System Be More Effective to Counteract Visual Bias Influenced by Topic Familiarity and Cognitive Tendencies?} \label{subsec:quantitative_RQ2}

Unlike \textit{H1} and \textit{H2}, which refer to the measures at the participant level, \textit{H3} and \textit{H4} require comparison at the trial level as each observation accounts for a different topic. Therefore, we re-ran the Shapiro–Wilk normality tests on the original 225 valid trial-level observations. The non-confirmatory visual attention ratio passed the normality test both for fixation counts (\(W = 0.9898, p = .6580\)) and fixation duration (\(W = 0.9748, p = .0540\)). Because the attention ratios are bounded between 0-1 and serve as the primary dependent variable in both hypotheses, we employed a beta regression model to examine whether individual factors (i.e., topic familiarity and cognitive tendencies towards bias) moderate the effect of interface condition on the attention ratio. The model specification is:
\text{Non-confirmatory\_attention\_ratio} (\text{duration} / \text{counts}) $ \sim $ \text{Factors} (\text{topic\_familiarity} / \text{cognitive\_tendencies}) $\times$ \text{interface} + (1 $\mid$ \text{user\_id}) + (1 $\mid$ \text{topic\_id}).

\subsubsection{Higher Topic Familiarity Did Not Significantly Predict Confirmatory Visual Attention or Moderate the Effect of the Debate System} \label{quant:familiarity} 

\textbf{Hypothesis Testing.} Regarding topic familiarity, as shown in Table \ref{tab:lme_results_familiarity}, we did not observe any significant main effect of the attention ratio as topic familiarity increased. The effect estimates were positive, the magnitude was small and non-significant ($\beta = +0.0205$, $p = .619$, for duration; $\beta = +0.0380$, $p = .341$, for counts). Additionally, the marginal effects were similarly modest ($+0.0012$ for duration, $+0.0026$ for counts). This suggests that the first half of H3 is not supported: \textbf{users with higher topic familiarity did not exhibit a significantly lower non-confirmatory visual attention ratio.} In other words, participants who were more familiar with the topic did not choose to read more attitude-consistent content.


\begin{table}[ht]
\centering
\resizebox{\textwidth}{!}{%
\begin{tabular}{@{}p{3cm}cccccc@{}}
\toprule
\textbf{Factors} & 
\multicolumn{2}{c}{\textbf{Slope ($\beta$) [Proportion]}} & 
\multicolumn{2}{c}{\textbf{$p$}} & 
\multicolumn{2}{c}{\textbf{Avg. Marginal Effect ($AME$)}} \\
\cmidrule(lr){2-3} \cmidrule(lr){4-5} \cmidrule(lr){6-7}
& Duration & Counts & Duration & Counts & Duration & Counts \\
\midrule
(Intercept) & 
$+0.1113$ [0.5280] & $+0.0626$ [0.5156] & 
$.401$ & $.626$ & 
-- & -- \\
Debate & 
$+0.0276$ [0.5069] & $+0.1046$ [0.5261] & 
$.887$ & $.582$ & 
$-0.0202$ & $-0.0200$ \\
Topic Familiarity & 
$+0.0205$ [0.5051] & $+0.0380$ [0.5095] & 
$.619$ & $.341$ & 
$+0.0012$ & $+0.0026$ \\
Interaction & 
$-0.0366$ [0.4909] & $-0.0622$ [0.4845] & 
$.554$ & $.304$ & 
$-0.0091$ & $-0.0155$ \\
\bottomrule
\end{tabular}
}
\begin{flushleft}
\small
Proportions in brackets are back-transformed from logit coefficients using the logistic function. The value is interpreted as an estimated non-confirmatory visual attention ratio. Marginal effects represent the average change in predicted probability (in percentage points) when a predictor variable increases by one unit.
\end{flushleft}
\caption{Beta mixed-effects regression model results for non-confirmatory visual attention ratio, with topic familiarity and interface condition as predictors. No significant effects were found for either the interface conditions or topic familiarity on visual attention to attitude-inconsistent content, with all effects showing non-significant p-values and minimal marginal effects.}
\label{tab:lme_results_familiarity}
\end{table}


The interaction between topic familiarity and the debate system yielded slightly negative coefficients ($\beta = -0.0366$, $p = .554$, for duration; $\beta = -0.0622$, $p = .304$, for counts), corresponding to proportions of $0.4909$ and $0.4845$, respectively. While not reaching traditional statistical significance, the negative coefficients mean the debate system became less effective in promoting non-confirmatory visual attention as topic familiarity increased. The marginal effects for this interaction were also negative ($-0.0091$ for duration and $-0.0155$ for counts). Additionally, the overall effect size (pseudo-$R^2$) was small for both fixation duration ($pseudo$-$R^2 = 0.008$) and fixation counts ($pseudo$-$R^2 = 0.012$), indicating that interface condition and topic familiarity together explained little variance in non-confirmatory visual attention. Together, this indicated that \textbf{no evidence support that topic familiarity significantly moderated the effect of the debate system on non-confirmatory visual attention.} \textbf{Overall, these results did not support H3.} 

\textbf{Qualitative Results.} From the interviews, we identified several factors that may explain why participant topic familiarity did not significantly affect their visual attention patterns. First, participants with lower topic familiarity reported naturally adopting balanced information-seeking strategies across both systems. P33 noted that ``\textit{on the topics where I wasn't as strong, I would look at both sides and maybe focus more on the opposite side,}'' and P26 similarly explained that ``\textit{for the topics I was less familiar with, I chose a broad mix, just so I could get an unbiased perspective, see both ends of the spectrum, and weigh everything.}''

Second, \textbf{participants who were more familiar with a topic appeared to hold higher expectations for argument quality and depth in the debate system}, and disengaged when these expectations were not met. For example, P35 reported that ``\textit{I thought that would provide some interesting insights and help me understand more reasons for supporting either one side or the other}'' and P28 noted that ``\textit{I already understood that there were multiple arguments on both sides so I seek new information to decide.}'' Together, these patterns offer a potential explanation for why the debate system became less effective at promoting non-confirmatory visual attention as topic familiarity increased.

\subsubsection{Higher Cognitive Tendencies Significantly Predicted Confirmatory Visual Attention, and the Debate System Provided a Marginal Buffering Effect.} \label{quant:tendency}


\textbf{Hypothesis Testing.} Considering the influence of individual cognitive tendencies, as shown in Table \ref{tab:lme_results_bias}, we observed a significant main effect: as tendency scores increased, the attention ratio decreased (duration: $\beta = -0.1315$, $p < .05$; count: $\beta = -0.1171$, $p < .05$), with estimated proportion of 0.467 and 0.471, respectively. The marginal effects for this predictor were $-0.0171$ for duration and $-0.0181$ for count, suggesting that each unit increase in cognitive tendencies reduced the non-confirmatory visual attention ratio by approximately 1.7--1.8 percentage points. This supported the first-half of H4: \textbf{users with higher cognitive tendencies towards bias exhibited less non-confirmatory visual attention ratio.} Simply put, participants with higher cognitive tendencies towards bias do read more attitude-consistent content. This result is different from participants with higher topic familiarity (Section \ref{quant:familiarity}).

\begin{table}[ht]
\centering
\resizebox{\textwidth}{!}{%
\begin{tabular}{@{}p{3.4cm}cccccc@{}}
\toprule
\textbf{Factors} & 
\multicolumn{2}{c}{\textbf{Slope ($\beta$) [Proportion]}} & 
\multicolumn{2}{c}{\textbf{$p$}} & 
\multicolumn{2}{c}{\textbf{Avg. Marginal Effect ($AME$)}} \\
\cmidrule(lr){2-3} \cmidrule(lr){4-5} \cmidrule(lr){6-7}
& Duration & Counts & Duration & Counts & Duration & Counts \\
\midrule
(Intercept) & 
$+0.4352$ [0.6071] & $+0.4113$ [0.6014] & 
$< .001$ & $< .001$ & 
-- & -- \\
Debate & 
$-0.3450$ [0.4146] & $-0.2973$ [0.4262] & 
$.038$* & $.069$ & 
$-0.0201$ & $-0.0201$ \\
Cognitive Tendencies & 
$-0.1315$ [0.4672] & $-0.1171$ [0.4708] & 
$.011$* & $.020$* & 
$-0.0171$ & $-0.0181$ \\
Interaction & 
$+0.1323$ [0.5330] & $+0.1086$ [0.5271] & 
$.085$ & $.150$ & 
$+0.0322$ & $+0.0265$ \\
\bottomrule
\end{tabular}
}
\begin{flushleft}
\small
Proportions in brackets are back-transformed from logit coefficients using the logistic function. It is interpreted as an estimated non-confirmatory visual attention ratio. Marginal effects represent the average change in predicted probability (in percentage points) when a predictor variable increases by one unit.
\end{flushleft}
\caption{Beta mixed-effects regression model results for non-confirmatory visual attention ratio, with cognitive tendencies and interface condition as predictors. Significant effects were found when switching from the baseline system to the debate system, showing a reduction in attention ratio in the absence of cognitive tendencies towards bias. Note that this analysis maps out the condition of when cognitive tendencies equal zero, which lies outside the observed sample range (0.28--0.75). Thus, we interpreted it as a mathematical baseline. Significant effects also existed as cognitive tendencies increased, leading to decreased attention ration in the absence of system changes. The interaction effect is positive but does not reach statistical significance (p = .085 for duration; p = .150 for count); it is consistent with a possible buffering effect but should be interpreted with caution rather than as a confirmed result.}
\label{tab:lme_results_bias}
\end{table}

The interaction between system and cognitive tendencies showed a theoretically meaningful pattern, though statistical significance varied by fixation measure. For fixation duration, the interaction coefficient was $\beta = +0.1323$ ($p = .085$, marginally significant), and for fixation count, $\beta = +0.1086$ ($p = .150$), with corresponding proportions of 0.533 and 0.527, respectively. The average marginal effects for the interaction were +0.0322 for duration and +0.0265 for counts. Specifically, this interaction helped increase their non-confirmatory visual attention ratio by approximately 2.6-3.2 percentage points. Additionally, the overall model pseudo-$R^2$ was small but larger than topic familiarity, for both fixation duration ($pseudo$-$R^2 = 0.034$) and fixation counts ($pseudo$-$R^2 = 0.030$), suggesting that cognitive tendencies and interface condition together account for a modest but meaningful portion of variance in non-confirmatory visual attention compared to topic familiarity. Although these estimates are positive and point in the direction of a buffering effect (the debate system may attenuate confirmatory attention among participants with stronger cognitive tendencies), the effect is not statistically significant and is present only for the duration measure at the marginal level. We therefore interpret it as a suggestive but inconclusive pattern rather than an established effect. Any benefit of the debate system for these participants should thus be regarded as tentative and in need of confirmation in adequately powered studies. \textbf{Taken together, H4 was not statistically supported, although the marginally significant duration interaction was directionally consistent with the hypothesized buffering pattern}. 

To further examine whether the two moderators (topic familiarity and cognitive tendency) operate jointly, we conducted an exploratory three-way interaction analysis (\ref{appendix:three-way}). The three-way interaction was non-significant for both fixation duration ($\beta = 0.012$, $p = .880$) and fixation counts ($\beta = 0.052$, $p = .497$), suggesting that topic familiarity and cognitive tendencies independently rather than jointly influence non-confirmatory visual attention.

\textbf{Qualitative Results.} From the interviews, we found that directly identifying participants' cognitive tendencies is challenging, as participants rarely articulate their own susceptibility to bias. One exception was P3, who explicitly acknowledged their confirmatory tendency: ``\textit{My original thought is to find something that advocated for what I'm already thinking.}'' P3 then noted that the debate system encouraged them to seek out opposing personas in hopes of changing their mind, suggesting that the system's structure helped counteract this tendency.

More broadly, participants' positive reflections on the debate system offer a potential explanation for why it partially benefited those with stronger cognitive tendencies. First, participants found the persona-driven arguments \textbf{more relatable and emotionally engaging} than the fact-based retrieval system, which may have lowered resistance to engaging with opposing viewpoints. P21 noted that ``\textit{it was easier to go through the debate version because I could connect with it more; it offers a more personal, human touch to the discussion.}'' Second, the diversity of stakeholder roles simulated by the personas helped \textbf{ground abstract controversial topics in everyday realities, making opposing perspectives feel more accessible}. As P2 explained, ``\textit{The truth, the science, and the facts are more for the professional. But it's important to include individual opinions; we should know how people from different backgrounds interpret them.}'' This sense of equitable representation was further reflected in how participants responded to inconsistencies in persona arguments. For example, P31 noted that a simulated cosmetics representative sometimes gave contradictory arguments about animal testing. However, they did not view this as a system flaw. Instead, P31 interpreted it as reflective of how real people reason through trade-offs: ``\textit{Not necessarily like what they're [personas] saying was wrong, that's what we [real people] would actually say, and sometimes we contradict ourselves.}''

Together, these qualitative insights suggest that the debate system's human-like and relatable presentation may be the key mechanism through which it partially reduced confirmatory attention among participants with stronger cognitive tendencies.

\section{Discussion} \label{sec:discussions}

\subsection{Summary of Mixed-Method Findings}

Through a mixed-method approach, we gain a multifaceted understanding of the debate system's effectiveness in mitigating confirmation bias. Based on prior work (Section \ref{subsec:llm simulation}), we expected the debate system to substantially mitigate confirmation bias. Instead, the system's effectiveness may depend on individual differences, benefiting primarily those with stronger cognitive tendencies toward bias. \textbf{This finding serves as a reality check: the multi-persona debate system is not necessarily superior to traditional fact-based retrieval that presents opposing viewpoints. Researchers should therefore interpret claims about the debiasing benefits of such systems with caution}.

\begin{table}[ht]
\centering
\small
\resizebox{\textwidth}{!}{%
\begin{tabular}{@{}p{4.1cm}p{7cm}p{7cm}@{}}
\toprule
\textbf{Research Questions} & \textbf{Hypothesis testing} & \textbf{Qualitative results} \\
\midrule

RQ1: Visual Attention (H1) &
Both systems yielded non-confirmatory ratios $> 0.5$; no significant difference between systems ($t = -0.961$, $p = .343$, $d = -0.158$). H1 is not supported &
Participants reported deliberately seeking balanced perspectives in both systems, regardless of system type. \\

\addlinespace

RQ1: Belief Change (H2) &
Positive belief change in both systems; no significant difference ($t = -0.288$, $p = .775$, $d = -0.047$); interaction between attention and debate system was negative. H2 is not supported &
Retrieval system perceived as trustworthy but dense and difficult to process; debate system arguments perceived as relatable but repetitive and lacking depth. \\

\addlinespace

RQ2: Topic Familiarity (H3) & Topic familiarity did not significantly predict confirmatory attention or moderate the effect of the debate system ($\beta = -0.037$, $p = .554$, $pseudo$-$R^2 = 0.008$). H3 is not supported &
Less familiar participants naturally sought balanced viewpoints; more familiar participants disengaged when debate arguments failed to meet their expectations for depth and novelty. \\

\addlinespace

RQ2: Cognitive Tendencies toward Biases (H4) &
Cognitive tendencies significantly predicted confirmatory attention ($\beta = -0.131$, $p = .011$); debate system showed marginal buffering effect ($\beta = +0.132$, $p = .085$, $pseudo$-$R^2 = 0.034$). H4 is partially supported &
Participants acknowledged confirmatory tendencies; the debate system's human-like, relatable presentation lowered resistance to opposing views. \\

\bottomrule
\end{tabular}
}
\begin{flushleft}
\small
This joint display cross-tabulates key quantitative findings with supporting qualitative evidence for each hypothesis. Effect sizes reported as Cohen's $d$ for paired $t$-tests and $pseudo$-$R^2$ for beta regression models.
\end{flushleft}
\caption{Joint display of quantitative and qualitative findings across research questions.}
\label{tab:joint_display}
\end{table}

\subsubsection{The effectiveness of the Debate system in reducing confirmation bias (RQ1)}
Both debate and retrieval systems encouraged participants to engage with diverse opinions that could challenge their prior beliefs. This is supported by the quantitative results, finding that participant non-confirmatory visual attention ratios were above 0.5 and belief change scores were positive in both systems (Section~\ref{quant:oppose H1}). Notably, this pattern appears to be driven less by system design than by participants' own deliberate information-seeking strategies: actively selecting opposing personas or horizontally scanning pro and con content regardless of interface type (Section~\ref{quant:oppose H1}). This suggests that \textbf{when opposing viewpoints are made explicitly available, motivated users will seek them out irrespective of how they are presented}.

However, the debate system did not significantly outperform the retrieval system in increasing non-confirmatory visual attention (Section~\ref{quant:oppose H1}), nor did it produce greater belief change (Section~\ref{quant:oppose H2}). Participants who devoted more attention to debate content were slightly less likely to shift their beliefs (Section~\ref{quant:oppose H2}). This may be because prolonged engagement exposed participants to repeated arguments rather than substantially new evidence, limiting the system’s ability to change existing beliefs. The null results of H1 and H2 constitute a substantive finding: when opposing viewpoints are made explicitly available, both systems encourage comparable engagement with attitude-inconsistent content. This serves as a necessary reality check on overly optimistic claims about the advantages of multi-persona debate systems.

While participants in both systems generally devoted substantial attention to attitude-inconsistent content, additional analysis suggests that confirmation bias might operate earlier in the information-seeking process through persona selection bias. This highlights the importance of distinguishing between selection and attention mechanisms when evaluating the debate system's effectiveness against confirmation bias. Consequently, even when opposing viewpoints are made available, the quality and depth of the presented arguments may matter more than the presentation format in shaping individual belief.


\subsubsection{The effectiveness of the Debate system on countering biased visual attention influenced by individual topic familiarity and cognitive tendencies (RQ2)}

We observed a notable difference in how individuals' topic familiarity and cognitive tendencies influenced participant visual attention and how the debate system interacted with both factors.

\textit{Topic Familiarity.} 
Contrary to our expectations, higher topic familiarity did not significantly predict more confirmatory visual attention (Section~\ref{quant:familiarity}). Interestingly, more familiar participants were slightly more deliberate in seeking opposing viewpoints. When the debate system failed to deliver sufficiently novel or deep counterarguments, familiar participants disengaged (Section~\ref{quant:familiarity}). This indicates that \textbf{topic familiarity acts less as a direct source of confirmation bias and more as a factor shaping expectations about argument quality: users who already know the topic may evaluate content more critically}. We further discuss the underlying factors that drive this familiarity effect in Section~\ref{subsec:theory}.

\textit{Cognitive tendencies toward bias.} Unlike topic familiarity, our hypothesis testing revealed a statistically significant effect of cognitive tendencies: participants with a stronger cognitive predisposition paid more visual attention to attitude-consistent content (Section~\ref{quant:tendency}). The debate system provides a suggestive but inconclusive trend to mitigate this negative effect. Qualitative data suggests that its human-like, relatable presentation lowered resistance to engaging with opposing viewpoints among those most prone to confirmation bias. This finding carries an important implication: \textbf{the advantages of the debate system are not shared equally by all users; they tend to be stronger for individuals who exhibit pronounced cognitive tendencies toward bias}. However, given that this buffering effect did not approach the strict statistical significance threshold, these conclusions should be interpreted cautiously.

Taken together, the null result for H3 and the partial support for H4 indicate that individual differences remain important for understanding visual attention. Although topic familiarity did not significantly moderate the interface effect, cognitive tendencies significantly predicted visual attention across conditions. The debate system showed a suggestive buffering pattern for stronger cognitive tendencies, but this moderating effect was not conclusively established. This pattern, combined with the null results for H1 and H2, suggests that the effectiveness of multi-persona debate systems is conditional rather than universal. When designing and evaluating such systems as interventions against confirmation bias, researchers should pay closer attention to individual differences and identify for whom and under what conditions it works.

\subsection{Theoretical Implications} \label{subsec:theory}


Our findings suggest that when people explore controversial topics, confirmation bias is best manifested as the result of an interaction between individual cognitive tendencies and system design, rather than as a purely stable cognitive trait or a direct behavioral outcome motivated to support one's own prior beliefs. This adds new knowledge to how confirmation bias is traditionally conceptualized and studied. 

Prior research has examined confirmation bias from either a behavioral perspective, which interprets selective exposure and polarized engagement as observable signs of user biases \citep{Suzuki2021-yn, Vedejova2022-cq}, or a cognitive perspective, focusing on internal processing tendencies such as cognitive predispositions \citep{Wang2024-hg, berthet2021measurement}. Our results directly reveal the relationship between both: cognitive tendencies significantly predict attitude-consistent attention. Moreover, we also find that this relationship is not fixed or automatic. Whether cognitive tendencies translate into biased information behavior depends on how information is structured and presented. In particular, the positive interaction between the debate interface and cognitive tendencies was directionally consistent with a possible buffering effect, although the interaction was not conclusively significant. Additionally, topic familiarity does not amplify biased attention as we assumed and may even support more deliberate cross-attitudinal engagement. Together, these findings extend prior work by demonstrating that the pathway of confirmation bias from cognition to behavior is context-sensitive.


To effectively use debate systems for promoting critical thinking and balanced information-seeking, our findings point to two fundamental user behavior tradeoffs that require careful consideration for future system design.


\subsubsection{Tradeoffs between dense, factual information processing and persuasive engagement} \label{subsec:tradeoffs-1} 

From the perspective of information foraging theory \citep{Pirolli1999-vi}, the divergent information presented in both systems can be perceived as information scent that attracts users seeking valuable information. Such information may serve different needs, including facilitating critical learning or providing counterarguments that users can use to reinforce existing beliefs. This interpretation is consistent with our finding that both systems produced non-confirmatory visual attention ratios exceeding 0.5, indicating that participants actively attended to opposing viewpoints regardless of interface format. 

However, increased attention to opposing information did not translate into stronger belief change, as reflected in the unsupported H2. Cognitive load theory \citep{Sweller2011-je} provides one explanation for this pattern. In the baseline system, the presentation of dense, factual information likely imposed substantial cognitive demands, requiring participants to devote considerable effort to understanding the content. Interview data suggest that this information density often created confusion rather than persuasive evidence capable of motivating belief change. By contrast, the debate system's persona-based conversational format may have reduced the effort required to process information, as reflected in lower visual attention compared with the baseline condition. However, this reduction in cognitive effort may have come at the cost of argumentative depth and evidentiary rigor, limiting the system's persuasive effectiveness through a debate mechanism. Together, these findings suggest a fundamental tradeoff between factual rigor and persuasive engagement: retrieval systems provide richer evidence but require greater cognitive effort, whereas debate systems facilitate engagement but may offer less compelling support for belief revision.

Importantly, participants did not experience this tradeoff uniformly. Our findings on topic familiarity suggest that users approached the systems with different epistemic goals. Participants with greater familiarity may have placed greater value on factual accuracy, evidentiary quality, and the critical evaluation of competing claims. For these individuals, dense, source-linked retrieval systems may have better matched their evaluative needs despite the higher cognitive demands involved. By contrast, participants with lower familiarity may have prioritized understanding and engagement, making conversational explanations that were easier to process and more relatable, particularly valuable. In such cases, persuasive engagement may facilitate continued information processing. Topic familiarity therefore appears to shape how users navigate the tradeoff between engagement and evidentiary rigor. The qualitative reports of disengagement among some more familiar participants may thus reflect a mismatch between the system conversational strengths and the needs of users seeking deeper evidence-based reasoning, rather than an inherent limitation of the debate approach.

These findings highlight the first design challenge: reducing the cognitive effort required to process factual information without sacrificing its rigor and persuasive potential. Future systems may address this challenge by translating technical language and statistical evidence into more accessible forms, progressively disclosing supporting details, or adaptively tailoring presentation styles to user familiarity and epistemic goals. Such approaches could support meaningful engagement with diverse perspectives while preserving the depth and credibility necessary for informed belief evaluation. We elaborate on these practical design implications in Section~\ref{design-implications-1}.

\subsubsection{Tradeoffs between diverse viewpoints exploration and convergent, actionable solutions} \label{subsec:tradeoffs-2} 

On one hand, participants valued the debate system for presenting nuanced and sometimes inconsistent persona reasoning, which encouraged critical reflection and careful evaluation of competing arguments. Returning to information foraging theory \citep{Pirolli1999-vi}, divergent viewpoints function as high-scent patches that users forage when they perceive sufficient informational value in exploring competing perspectives. However, this informational value diminishes when content becomes cognitively demanding or when prolonged engagement induces fatigue. Users shift toward seeking convergent, decision-ready information, consistent with the patch-leaving prediction of optimal foraging models. On the other hand, participants expressed a desire for clearer, more concrete recommendations to guide decision-making. While convergent answers can facilitate choices, they also carry the risk of automation bias, leading users to accept system suggestions uncritically \citep{Nguyen2018-tu, Bucinca2021-uc, Lai2023-bu}. 

Taken together, these implications highlight the need for future research to carefully determine when debate systems should prioritize divergent perspectives versus providing convergent solutions. In Section~\ref{design-implications-2}, we present practical design recommendations for navigating this tradeoff.

\subsection{Practical Implications}

\subsubsection{Integrating user-adaptive text simplification with persona-based reframing} \label{design-implications-1}

To enhance the readability of factual content while maintaining the human-like, vivid delivery of LLM-generated multi-persona systems, both their technical implementation and feature design need refinement. We suggest that future multi-persona debate systems should translate scientific content into easy-to-read arguments and reflect diverse stakeholders' viewpoints. Achieving this requires two complementary mechanisms: improving readability through text simplification and enhancing argument resonance through persona-based reframing. Grounded in prior work, we outline a three-step integration pipeline below.

\textbf{Step 1: User familiarity profiling.} Future debate systems could first use participants' self-reported topic familiarity scores (e.g., already collected as part of pre-task questionnaires in our study) as system prompts to determine the appropriate level of content complexity before argument generation. \citet{Kim2025-wr} showed that AI systems can apply user knowledge levels to generate scientific explanations that better align with their comprehension needs.

\textbf{Step 2: Audience-adaptive text simplification.} Based on the profiled familiarity level, the system selects an appropriate simplification strategy before generating each persona argument. Recent work has demonstrated that LLMs can achieve targeted concept simplification for domain-specific texts by instructing the model to reframe technical content for a specified audience level while preserving factual accuracy \citep{Asthana2024-oq}. For users with lower topic familiarity, this could range from a simple plain language instruction \citep{Arias-Russi2025-dh} to a more structured approach using rhetorical planning frameworks that organize content around discourse goals and audience comprehension needs \citep{Liu2025-iq}. However, one caveat is that what counts as ``simplified'' is often subjective \citep{Chen2025-ue}, which motivates the need for a quality verification step.

\textbf{Step 3: Readability verification and persona-based reframing.} Before display, generated arguments could be evaluated on quality dimensions such as clarity, coherence, and vocabulary complexity using an LLM-as-judge approach. \citet{Belem2025-dg} have found that this reference-free evaluation strategy aligned well with human-annotated rationales. Arguments passing this quality filter are then reframed through the voice of simulated personas reflecting specific stakeholder perspectives.

Integrating user-adaptive simplification with persona-based reframing across these three steps may offer a more robust design strategy for aligning scientific content with diverse audiences while preserving factual integrity.

\subsubsection{Scaffolding the divergence and convergence in multi-persona debate} \label{design-implications-2}

Building on existing technical constraints of LLM-generated multi-persona debate, we describe a three-step scaffolding approach for managing the transition between divergent and convergent phases of debate. Grounded in prior work and our system design, we outline this integration below.

\textbf{Step 1: Detecting debate convergence.} A key challenge in sustaining productive debate is identifying when personas begin to converge or repeat arguments without generating new insights. \citet{Liang2023-ml} refer to this as the ``degeneration-of-thought'' problem, in which LLMs stop producing new ideas once they reach sufficient confidence in their positions. To address this, future systems could employ two complementary detection mechanisms. First, the system could automatically monitor semantic similarity between successive persona arguments, flagging repetition when similarity exceeds a defined threshold. Second, inspired by the framework proposed by \citet{Koupaee2025-mh}, which also uses multi-round debate to surface disagreements in a manner similar like our approach, future system design could add a dedicated adjudication stage in which an additional LLM agent evaluates whether remaining disagreements are genuinely unresolved or have converged to a point where further debate is unlikely to generate new perspectives.

\textbf{Step 2: Scaffolding divergence and convergence.} Once convergence is detected, the system faces a design choice: sustain divergent exploration or shift toward convergent solution finding. Drawing on the \textit{Rehearsal} system introduced by \citet{Shaikh2024-vs}, which scaffolds structured conflict simulation to facilitate resolution, future debate systems could adopt a similar staged approach. It first maximizes viewpoint diversity through persona disagreement, then progressively guides personas toward identifying shared ground. When convergence is detected, the system could trigger one of two responses: generate a new persona with a distinct stakeholder perspective to reintroduce divergence, or prompt existing personas to explicitly address unresolved disagreements and propose convergent solutions. To distinguish genuine consensus from premature agreement, future systems could adopt the confidence-weighted voting mechanism proposed by \citet{Chen2023-iv} in \textit{ReConcile}. Each persona provides a confidence score alongside its argument, and convergence is accepted only when the weighted consensus surpasses a predefined threshold. This prevents the system from collapsing into superficial agreement while meaningful disagreement remains.

\textbf{Step 3: Reliability indicators for debate content.} A critical concern in persona-based debate is the risk of presenting unsupported or misleading arguments as equally credible to evidence-backed claims. This risk is particularly acute for topics with established scientific consensus, where presenting opposing arguments may inadvertently amplify misinformation \citep{Phillips2018-sm, Procter2023-rr}. To address this, future systems could accompany each persona argument with a reliability indicator. For example, distinguishing between evidence-backed claims, reasoned opinion, and unverified assertions enables users to critically evaluate argument quality as they engage with the debate. This categorization can be effectively achieved using conventional claim detection systems \citep{Hassan2017-fn} or by adopting an LLM-as-judge approach \citep{Ni2024-yi}. For topics with strong scientific consensus, the system could additionally impose evidentiary thresholds, filtering out arguments that fail to meet minimum reliability standards and thereby preventing the debate from misrepresenting settled empirical questions as open controversies.

\subsection{Limitations} \label{subsec:limitations}

\subsubsection{System design and development} \label{subsec:limit-design}

Our current multi-persona debate system has several design limitations that may have influenced the results. We organize these into three areas: user engagement and argument variability, LLM factual grounding and inherent bias, and baseline comparisons.

\textbf{User engagement and argument variability.} 
The current system lacks some interactive features to increase user engagement, such as the ability to nudge debate conversations by incorporating user feedback and new inputs. Additionally, system improvements are needed to enhance debate quality, such as refining prompting techniques to minimize repetitive responses from personas, and facilitating a natural closure in the debate (e.g., ``I have nothing new to add''). Additionally, we cannot fully rule out novelty effects as a contributing factor to the observed results. The unfamiliar appearance of the debate system may have attracted visual attention independent of argument content. Although each experimental block began with a training task specifically designed to familiarize participants with the assigned system before data collection (as described in Section \ref{subsec:task design}), residual novelty-driven attention cannot be entirely eliminated. Future studies could address this by including a longer familiarization phase or a longitudinal design that captures changes in engagement over repeated system use.

Because personas and their arguments were dynamically generated for each session, participants encountered different debate content even when exploring the same topical queries, introducing noise into the experimental conditions and limiting strict reproducibility. Additionally, because generated arguments also varied depending on participant persona selections, ranging from formal academic arguments to casual conversational responses, it is difficult to fully disentangle the effects of system design from argument type. To support replication efforts, we provide the full prompts used for persona generation and debate continuation in \ref{appendix:prompts}. Future studies could improve reproducibility by pre-generating and standardizing debate argument, and by systematically controlling persona types to better isolate the effect of content type on confirmation bias.

\textbf{LLM Factual grounding and inherent bias.} 
A notable design tradeoff concerns the decision to exclude retrieval-augmented generation (RAG) from the debate system. Although RAG was initially integrated to ground persona responses with external evidence, pilot testing revealed that it significantly increased response generation times, disrupting the natural flow of visual attention captured by our eye-tracking measures. While prioritizing experimental fluency was necessary for methodological consistency, this decision may have limited the factual rigor of persona arguments. Participants explicitly noted that persona arguments felt less factually grounded compared to the retrieval system's evidence-based content, which may have attenuated belief change even when participants attended to opposing viewpoints.

Future implementations should explore lightweight RAG approaches that minimize generation latency. For example, recent industrial applications have adopted simple inference-time factuality checkers \citep{Sankararaman2024-id, Park2026-wt} that determine whether to retrieve additional evidence on a per-claim basis, yielding a low-latency, low-cost solution that could be readily integrated into persona argument generation at run-time. This selective retrieval strategy would preserve argument quality and reduce hallucination risks without the response latency that disrupted eye-tracking measurement in our study, potentially strengthening the debate system's effectiveness in promoting engagement with opposing viewpoints.


Our implementation used GPT-4 as the primary backend, which was one of the most capable LLMs available at the time of the study and aligned with our university policy requirements. However, because the system has not been evaluated for potential biases in its representations across topics and personas, biases inherent to GPT-4 alignment training may have influenced our user testing results. While our system architecture is model-agnostic at the prompt and interface level (Section~\ref{subsec:system}), our internal assessment (\ref{appendix:llm-assessment}) evaluated GPT-4 against alternative configurations but was not exhaustive. Replication with other LLM backends would be valuable to assess whether debate content quality and bias-mitigating effectiveness generalize across models. Future studies should incorporate more robust techniques to detect and address potential model biases.


\textbf{Baseline generaliability.}
Finally, our baseline condition was built on top of ArgumentSearch rather than traditional or commercial search engines, which may limit the generalizability of our findings to more typical search scenarios. Comparative evaluations with broader baseline systems would strengthen the external validity of future work.

\subsubsection{Measurement Confounds} \label{subsec:measurement_confounds}

Reflecting on the hypothesis testing results, we identified several potential measurement confounds that were not considered during hypothesis formulation (Figure \ref{fig:procedure_diagram}). In this section, we elaborate on these issues to guide future researchers in taking additional steps.

\textbf{Topic familiarity as a proxy measure.} Our measure of topic familiarity uses a single-item, self-report scale measurement before each trial. This metric reflects participants' perceived familiarity but does not capture the underlying information-seeking intentions of individuals with different levels of familiarity. As discussed in Section~\ref{subsec:tradeoffs-1}, the reduced effectiveness of the debate system at higher levels of familiarity could potentially derive from a misalignment in epistemic goals. Participants with higher familiarity may be more oriented toward accuracy-focused information, whereas less familiar participants may prioritize information that is relatable and easy to understand. Accordingly, we propose that future research should examine the relationship between topic familiarity and different information-seeking intentions. This should make it easier to distinguish how much the debate system and the retrieval system each contribute to the overall effectiveness.

\textbf{Visual attention as a behavioral manifestation.} The non-confirmatory visual attention ratio (defined in Equation~\ref{eq:visual_attention_ratio}) captures the \textit{allocation} of fixation activity toward attitude-inconsistent content but cannot distinguish among the different motivations that may drive increased attention. In the context of confirmation bias, increased attention on opposing viewpoints could reflect different purposes, such as genuine openness to attitude-inconsistent information \citep{Lu2025-hw}, active counterarguing in which users attend to opposing content in order to refute it, or conflict monitoring in which attitudinal inconsistency is detected but not deeply processed \citep{Frey1986-jp}. Our measure operationalizes reduced confirmatory visual attention as a behavioral indicator, without objectively revealing which of the above underlying processes are at play. Qualitative findings suggest that at least some elevated attention reflects genuine openness as exploratory information-seeking. Therefore, we encourage future work to incorporate more nuanced measures that capture these different intents and examine how they relate to patterns of visual attention. Specifically, we could ask whether visual attention corresponds proportionally to particular intents, and whether these intents evolve over time as individuals continue engaging with diverse information.

\subsubsection{Sampling, test time, and topical differences} \label{subsec:limit-procedure}
We also recognize that our findings are limited by the participant pool, which primarily consisted of college students and was politically skewed toward liberal viewpoints. This homogeneity limits the external validity of our results in several key respects. First, college students are likely to have greater familiarity with digital technologies and LLM-based systems than the broader public, which may either exaggerate or diminish observed engagement effects. Second, the age range and educational attainment of our participants do not reflect other segments of the population, such as older adults, individuals with lower digital literacy, or non-English-speaking communities, all of whom may engage with controversial information and debate systems in distinct ways. Third, our sample showed an uneven political ideological distribution, with most participants tending to be liberal. Because political ideology can affect how individuals interpret controversial topics and judge opposing perspectives, this skew may have shaped their evaluations of the debate system, limiting the generalizability of our findings across the political spectrum. Finally, the visual criteria required for eye-tracking participation further exclude users in this age group who have certain accessibility-related needs. 

Beyond sample composition, our study limited participants' time in the lab, which may have limited their ability to fully explore long-term interaction patterns with the system. Extended use of the system could potentially reveal different engagement behaviors and belief changes that were not captured within the current study time constraints.

Additionally, although our study intentionally selected diverse topics across different domains to balance the limitations of single-topic studies with the goal of cross-topic generalizability, our topic-sensitivity analysis (Section \ref{appendix:topic-sensitivity}) reveals nuanced differences at the individual topic level, such as animal testing, and vaping e-cigarettes. Either visual attention or belief change of these topics slightly diverges from the overall trends. This variation suggests that inherent topic characteristics may influence the direction of belief change in ways not fully captured by aggregate-level analyses, warranting further investigation in future work.

Future research should therefore consider both broader participant samples and longer engagement periods to develop a more comprehensive understanding of how debate systems function across varied user groups. In particular, it would be valuable to examine how age and digital or LLM literacy shape users' willingness and ability to engage with persona-based systems. Recent work suggests that older adults tend to place less trust in LLM-based tools than in traditional search systems, citing concerns such as the opaque nature of LLM information retrieval \citep{Enam2025-hn} and difficulty managing extended conversational interactions \citep{Zubatiy2023-rb}. Analogous disparities related to education and cultural background may emerge in cross-cultural contexts \citep{Scherr2025-xe}, and warrant dedicated investigation.


\section{Conclusion} \label{sec:conclusion}

This study introduces and evaluates an LLM-powered multi-persona debate system designed to mitigate confirmation bias in information access.

The system presents diverse viewpoints through an interactive, persona-based interface where users engage with simulated debates across multiple perspectives on controversial topics. Using a within-subject, mixed-methods study design, we found that, compared to an opposing-stances retrieval system, the multi-persona debate system did not significantly increase users' visual attention to attitude-inconsistent content overall. Stronger cognitive tendencies toward bias were significantly associated with greater attention to attitude-consistent content across conditions. The debate system showed a positive buffering pattern among participants with stronger cognitive tendencies; this interaction was marginally significant for fixation duration but not significant for fixation count, and should therefore be regarded as suggestive rather than conclusive.

Our results offer valuable insights while providing a cautious assessment of the effectiveness of multi-persona debate systems in mitigating confirmation bias. 


\section{Acknowledgment}
This research was supported in part by several generous sponsors: by Microsoft, by Good
Systems (a UT Austin Grand Challenge dedicated to developing responsible AI technologies), by US National Science Foundation grant (\#2421782) and Simons grant (MPS-AI-00010515) awarded to the NSF-Simons AI Institute for Cosmic Origins (CosmicAI), and by the UT Austin iSchool, and by free computing credits from Amazon Bedrock and Google. We extend our special thanks to Gauri Kambhatla for her assistance
regarding polarization research, and to Anubrata Das and Sooyong Lee for their valuable
feedback and insights. We are also grateful to our participants, without whom this research
would not have been possible. The statements made herein are solely the opinions of the authors
and do not reflect the views of the sponsoring agencies.


\printbibliography

\appendix
\section*{Appendix} 

\section{Measurements for Data Collection} \label{appendix:measurement}
\subsection{Pre-task questionnaire}
\begin{enumerate}
    \item Please indicate your current stance on the given topic? [\textit{Supportive; Neutral/ Undecided; Opposed}]
    \item How confident are you in your stance and reason? [\textit{Very low; Slightly low; Moderate; Slightly high; Very high}]
    \item How familiar are you with the topic? [\textit{Not familiar; Slightly familiar; Moderately familiar; Familiar; Very familiar}]
\end{enumerate}

\subsection{Post-task questionnaire (measuring topic familiarity)} 
\begin{enumerate}
    \item Please indicate your current stance on the given topic? [\textit{Supportive; Neutral/ Undecided; Opposed}]
    \item How confident are you in your stance and reason? [\textit{Very low; Slightly low; Moderate; Slightly high; Very high}]
\end{enumerate}

\subsection{Post-system questionnaire (measuring individual cognitive tendencies toward bias)} \label{app:cb_survey}

This questionnaire is adopted from \cite{berthet2021measurement}. Participants were instructed as follows: \textit{``In the following questionnaire, you will be asked to evaluate a candidate's personality trait. Please read the description of the personality first, then select 8 questions from the 20 listed questions to evaluate if the candidate has the given personality trait or not.''}

\textbf{Personality trait 1:} \textit{``This type of person is typically good-natured, gentle, and cooperative. They are generous, warmhearted, and do not mind giving of their time and energy. They are not prone to jealousy. They will accept blame if things go wrong, rather than putting the blame on others. They are willing to compromise and find ways of cooperating with others, even in difficult situations.''}

Please make 8 selections among these 20 listed questions to evaluate if a candidate has the given personality trait or not:
\begin{enumerate}
    \item Tell me about a time you held yourself publicly accountable for a mistake you made.
    \item What do you do to keep yourself in good spirits?
    \item What is it about working with other people that you enjoy most?
    \item What do you do to keep from resenting another person who seems to be getting all the breaks?
    \item Tell me about a situation in which you spent time listening to a person who wanted to talk about a problem, even though you had other important things to do.
    \item How has willingness to compromise contributed to your success?
    \item How do you maintain a positive outlook on life?
    \item Tell me about a situation in which cooperating with others really paid off for you.
    \item What kinds of people seem to rub you the wrong way?
    \item Tell me about a time when your strong opinions got you in trouble.
    \item Tell me about a time when your quick temper got you in trouble.
    \item What most irritates you about others?
    \item How does your jealousy of others usually show itself?
    \item What are some ways in which you've been able to use others to help you achieve your own ends?
    \item When was the last time when you tried to ``get even'' because someone did something that irritated you?
    \item Describe a situation in which you are likely to be argumentative rather than cooperative.
    \item What do you see as your greatest strengths? Your greatest weaknesses?
    \item What do you think are the most pressing problems facing America today?
    \item What magazines do you subscribe to?
    \item What would you say are the characteristics of a good teacher?
\end{enumerate}

\textbf{Personality trait 2:} \textit{``This type of person is typically responsible, orderly, and persevering. They can be depended on to accept responsibilities. They do not allow personal benefit or gain to conflict with ethical principles. They are painstaking and thorough, and see a job through in spite of difficulties or temptations. They are hardworking, strong-willed, and sometimes get overinvolved with details.''}

Please make 8 selections among these 20 listed questions to evaluate if a candidate has the given personality trait or not:
\begin{enumerate}
    \item Tell me about a situation where your concern for detail made it difficult to get a project done on time.
    \item Why is it that others feel they can depend on you?
    \item Describe a situation in which you avoided compromising your personal standards even under pressure to do so.
    \item What do you do to keep yourself motivated to finish a task on which you are losing interest?
    \item When was the last time you got so caught up in something you were doing that you forgot about everything else?
    \item Tell me about some way in which you feel you differ from the typical college student.
    \item Tell me about a situation where perseverance paid off for you.
    \item How did you feel the last time you resisted the temptation to take ``the easy way out''?
    \item How did you feel the last time someone depended on you to do something and you let them down?
    \item How often do you cut classes?
    \item When does your interest in your own personal life get in the way of getting important things done?
    \item Do you think you'd do better in school if you were more organized? Why?
    \item Tell me about a project you started and never finished.
    \item In what kinds of situations are you willing to stretch your ethics to get something you want?
    \item What was your reaction the last time you got a low grade on a paper because of typing errors, spelling mistakes, or bad grammar?
    \item What are the projects you just can't get motivated to start?
    \item How would other people describe you?
    \item What do you do to keep up with current events?
    \item What do you see yourself doing five years from now?
    \item What movie that you saw in the last year impressed you most?
\end{enumerate}

\textbf{Personality trait 3:} \textit{``This type of person is typically poised, calm, and composed. They rarely seem to get upset or nervous. They go on with what they are doing regardless of distractions. They do not lose their composure when provoked or when in an emergency situation. They perform well under pressure.''}

Please make 8 selections among these 20 listed questions to evaluate if a candidate has the given personality trait or not:
\begin{enumerate}
    \item Have you ever been in a situation where you kept your cool while everyone else fell apart?
    \item Do people ever comment on how poised you appear when speaking in front of a group?
    \item Do you see yourself as more relaxed and easygoing than other people? Why?
    \item Tell me about a time when you handled a difficult situation well.
    \item When was the last time something unexpected came up which threw off your plans? How did you adapt to the situation?
    \item Tell me about a situation in which you felt yourself about to lose your cool. What did you do to prevent this from happening?
    \item Describe a situation in which you finished a project in spite of pressure and distractions.
    \item What do you do to keep calm in pressure situations?
    \item What is the greatest source of anxiety for you?
    \item What sort of situations embarrass you?
    \item Did you ever have any nervous habits which you have now overcome?
    \item Is it ever difficult to keep your nervousness from showing?
    \item What is it about pressure situations that prevents you from doing your best work?
    \item Do you find yourself snapping at other people over unimportant things?
    \item What kinds of social situations make you feel ill at ease?
    \item How do you show your nervousness?
    \item In a few sentences, how would you describe yourself?
    \item What course did you enjoy most in college? Why?
    \item What person has had the greatest influence on your life? Why?
    \item Describe the ideal job.
\end{enumerate}

\textbf{Personality trait 4:} \textit{``Extroverts are typically outgoing, sociable, energetic, confident, talkative, and enthusiastic. Generally confident and relaxed in social situations, this type of person rarely has trouble making conversation with others. This type of person makes friends quickly and easily and is usually able to make a favorable impression on others. This type of person is usually seen by others as characteristically warm and friendly.''}

Please make 8 selections among these 20 listed questions to evaluate if a candidate has the given personality trait or not:

\begin{enumerate}
    \item What events make you feel popular with people?
    \item What do you do to keep yourself in good spirits?
    \item What social activities (e.g., clubs, groups, fraternities or sororities) have you been active in over the years?
    \item What do you like about living situations in which there are always lots of people around?
    \item In what social situations are you most likely to be outgoing and friendly?
    \item What do you think are the good and bad points of acting friendly and open?
    \item In what social situations are you most likely to feel self-assured and confident in yourself?
    \item What would you do if you wanted to liven things up at a party?
    \item In what situations do you wish you could be more outgoing?
    \item Tell me about a time when you felt left out from some social group. How did you handle these feelings?
    \item What kind of events make you feel like being alone?
    \item What factors make it hard for you to really open up to people?
    \item Describe a type of social situation that invariably makes you feel ill at ease and awkward. What is it about such situations that makes you feel uncomfortable?
    \item Think about times when your shyness in social situations has made you come across as being aloof. Give me an example.
    \item What things do you dislike about loud parties?
    \item Think about a time when you really wanted to talk to someone, but just couldn't bring yourself to initiate conversation. What types of situations are most likely to make you feel this way?
    \item What activities do you really excel in?
    \item What kind of charities do you like to contribute to?
    \item What are some of your favorite books?
    \item What are your career goals?
\end{enumerate}

\subsection{Interview questions}
\begin{itemize}
    \item When you were using the retrieval-based system,
    \begin{enumerate}
        \item what kind of information interested you most?
        \item What document did you decide to read?
        \item What were you looking for in the document page?
        \item Which interface did you find most useful?
        \item Did you find the content trustworthy/ useful? Why?
    \end{enumerate}
    \item When you were using the multi-persona debate system,
    \begin{enumerate}
        \item How did you decide which persona to select to engage in the debate?
        \item How did you decide how to switch agents between debate rounds?
        \item Why did you change (not change) the personas between rounds?
        \item How did you decide which persona to change/ keep between rounds?
        \item How many rounds of debate did you usually read? Why did you want to proceed to the next round? When would you stop generating the next round?
        \item What kind of content interested you most/ helped you to explore the topic and make a decision?
        \item Did you find the bookmark function useful? Why?
        \item How did you decide which information to save in the bookmark?
        \item Did you find the LLM-generated information useful and trustworthy?
    \end{enumerate}
    \item In general, how did you use these two systems to explore the diverse perspectives for the given topics? Where did you find most useful for your exploration?
    \item For the two systems you used in this study, 
    \begin{enumerate}
        \item Were you satisfied with the experience when using the systems?
        \item Were you overloaded by information? 
        \item Was the design effective?
        \item which one do you prefer? 
    \end{enumerate}
    \item Do you think your prior knowledge would influence how you use the systems? 
    \begin{enumerate}
        \item When you were familiar with the topic, how would you use the systems?
        \item when you were unfamiliar with the topic, how would you use the systems?
    \end{enumerate}
    \item When exploring with the system, what factors/ information would make you shift your perspective?
\end{itemize}

\pagebreak
\section{Participant Demographics} \label{appendix:participant demographics}

\begin{table}[h!]
\centering
\small
\resizebox{\textwidth}{!}{%
\begin{tabular}{@{}llllll@{}}
\toprule
\textbf{ID} & \textbf{Gender} & \textbf{Age} & \textbf{Highest Education} & 
\textbf{Discipline} & \textbf{Political Ideology} \\
\midrule
1  & Female & 23 & Bachelor’s degree & Applied sciences & Liberal \\
2  & Female & 22 & High school or equivalent & Social sciences & Liberal \\
3  & Female & 24 & Bachelor’s degree & Engineering & Moderately liberal \\
4  & Male & 28 & Bachelor’s degree & Natural sciences & Centrist \\
5  & Female & 19 & High school or equivalent & Social sciences & Moderately liberal \\
6  & Female & 24 & Bachelor’s degree & Social sciences & Strongly liberal \\
7  & Female & 28 & Bachelor’s degree & Other & Centrist \\
8  & Female & 20 & High school or equivalent & Natural sciences & Strongly liberal \\
9  & Male & 20 & High school or equivalent & Natural sciences & Liberal \\
10 & Female & 29 & Bachelor’s degree & Social sciences & Liberal \\
11 & Male & 25 & Master’s degree & Applied sciences & Liberal \\
12 & Female & 25 & Master’s degree & Natural sciences & Liberal \\
13 & Male & 32 & Master’s degree & Natural sciences & Centrist \\
14 & Female & 18 & High school or equivalent & Social sciences & Centrist \\
15 & Female & 29 & Professional degree (MD, JD, etc.) & Natural sciences & Liberal \\
16 & Male & 18 & High school or equivalent & Natural sciences & Moderately conservative \\
17 & Male & 29 & Doctoral degree & Natural sciences & Moderately conservative \\
18 & Non-binary\slash{}third gender & 20 & High school or equivalent & Natural sciences & Liberal \\
19 & Male & 18 & High school or equivalent & Applied sciences & Strongly liberal \\
20 & Male & 19 & Bachelor’s degree & Engineering & Strongly liberal \\
21 & Female & 25 & Bachelor’s degree & Engineering & Moderately liberal \\
22 & Non-binary\slash{}third gender & 20 & High school or equivalent & Natural sciences & Centrist \\
23 & Male & 26 & Master’s degree & Information Science & Liberal \\
24 & Male & 19 & High school or equivalent & Engineering & Moderately liberal \\
25 & Female & 19 & High school or equivalent & Applied sciences & Liberal \\
26 & Male & 18 & High school or equivalent & Engineering (B.S.) & Centrist \\
27 & Male & 19 & High school or equivalent & Natural sciences & Strongly liberal \\
28 & Female & 20 & High school or equivalent & Advertising & Liberal \\
29 & Male & 20 & High school or equivalent & Applied sciences & Centrist \\
30 & Female & 34 & Master’s degree & Humanities & Strongly liberal \\
31 & Female & 20 & High school or equivalent & Social sciences & Liberal \\
32 & Male & 20 & High school or equivalent & Engineering & Liberal \\
33 & Male & 21 & High school or equivalent & Applied sciences & Moderately liberal \\
34 & Male & 19 & High school or equivalent & Applied sciences & Centrist \\
35 & Female & 25 & Master’s degree & Natural sciences & Strongly liberal \\
36 & Female & 26 & Master’s degree & Information Studies & Liberal \\
37 & Non-binary\slash{}third gender & 20 & High school or equivalent & Engineering & Liberal \\
38 & Male & 20 & High school or equivalent & Applied sciences & Moderately conservative \\
39 & Male & 32 & Bachelor’s degree & Computer Science & Liberal \\
40 & Female & 20 & High school or equivalent & Applied sciences & Liberal \\
\bottomrule
\end{tabular}%
}
\caption{Participant demographic information and political ideology.}
\label{tab:participant_info}
\end{table}

\pagebreak

\section{Prompts} \label{appendix:prompts}

\subsection{Prompt Placeholders}
Throughout our prompts, we use placeholders, which are automatically replaced by their corresponding value as defined in our initial configuration. For all our current versions of the system, these values are:
\begin{tcolorbox}[
    colback=gray!5,
    colframe=gray!50,
    fonttitle=\small\bfseries,
    coltitle=black,
    breakable,
    enhanced]
\small
\RaggedRight
\begin{enumerate}
    \item \textbf{[TOPIC]}: The user-input topic to focus persona generation and debate discussion around.
    \item \textbf{[NUM\_PERSONAS]}: Initial requested number of personas
    \item \textbf{[CURRENT\_PERSONAS]}: Current personas in JSON format
    \item \textbf{[NAME]}: Name of the persona
    \item \textbf{[DESC]}: Description of the persona
    \item \textbf{[HISTORY]}: Transcript of the debate in JSON format
    \item \textbf{[LIMITER]}: Sentence telling the LLM how long their response should be (ie, Limit your response to 5 sentences).
    \item \textbf{[RESPONSE\_LENGTH]}: Sentence limit instructing the LLM how long its response should be (e.g., ``Limit your response to 5 sentences'').
\end{enumerate}
\end{tcolorbox}

\subsection{Persona Generation}

\label{genpersona}
\begin{tcolorbox}[
    title={\textbf{Generating an initial set of personas}},
    colback=gray!5,
    colframe=gray!50,
    fonttitle=\small\bfseries,
    coltitle=black,
    breakable,
    enhanced
]

\small
\RaggedRight
Given the topic, [TOPIC], create a roundtable on the debate of different personas to show key perspectives on it. Output the personas as a list of JSON objects. Each JSON object should have the following structure:

{ "title": <title of the Persona>, \\
"description": <Brief Description of the Persona, only describing their background (rather than their stance)>, \\
"emoji": <Single emoji representation of the perspective the persona represents>}. 

Ensure that the output is formatted as a valid JSON. Please generate exactly [NUM\_PERSONAS] personas.
\end{tcolorbox}

\begin{tcolorbox}[
    title={\textbf{Adding persona}},
    colback=gray!5,
    colframe=gray!50,
    fonttitle=\small\bfseries,
    coltitle=black,
    breakable,
    enhanced
]

\small
\RaggedRight
Given the topic, [TOPIC], we are creating a roundtable debate of different personas to show key perspectives on it. Currently, we have personas as follows: [CURRENT\_PERSONAS] Please output exactly one additional persona. Your output should have the following JSON structure:

{"title": <title of the Persona>,\\
"description": <Brief Description of the Persona, only describing their background (rather than their stance)>,\\
"emoji": <Single emoji representation of the perspective the persona represents>}.\\

Ensure that the output is formatted as a valid JSON.
\end{tcolorbox}

\subsection{Debate Generation}

\begin{tcolorbox}[
    title={\textbf{Starting a debate}},
    colback=gray!5,
    colframe=gray!50,
    fonttitle=\small\bfseries,
    coltitle=black,
    breakable,
    enhanced
]

\small
\RaggedRight
You are in a roundtable debate on the topic [TOPIC].\\
You are [NAME], who is [DESC].\\
Please start the debate by concisely presenting your
argument for your stance on the topic. [LIMITER]
\end{tcolorbox}

\begin{tcolorbox}[
    title={\textbf{Continuing a debate}},
    colback=gray!5,
    colframe=gray!50,
    fonttitle=\small\bfseries,
    coltitle=black,
    breakable,
    enhanced
]

\small
\RaggedRight
You are in a continuing roundtable debate on the topic [TOPIC].\\
You are [NAME], who is [DESC].\\
Here is the transcript of the debate so far: [HISTORY]\\
Please continue to debate the others, concisely supporting your stance on
the topic. [LIMITER]
\end{tcolorbox}

\subsection{Generated Examples}

We present example debates generated on the topic: ``Should animals be used for scientific or commercial testing?'' across three persona positions: pro (supporting the question), con (opposing the question), and neutral (discussing both pro and con perspectives on the question).

\begin{tcolorbox}[
    title={\textbf{First round: Presenting personas, initial positions, and opening statements}},
    colback=gray!5,
    colframe=gray!50,
    fonttitle=\small\bfseries,
    coltitle=black,
    breakable,
    enhanced
]

\begin{tcolorbox}[
    title={\textbf{Animal Rights Advocate} \hfill \textcolor{red}{\textbf{CON}}},
    colback=red!5,
    colframe=red!40,
    fonttitle=\small\bfseries,
    breakable,
    enhanced
]
\small
\textit{A recognized activist who has dedicated their life to advocating for animal rights. They are a member of several international animal welfare organizations.}

\medskip
As a life-long advocate for animal rights, I firmly oppose the use of animals in scientific or commercial testing due to the following reasons:
\begin{itemize}
    \item \textbf{Ethical Concerns:} Animals are sentient creatures, capable of feeling pain, distress, and fear. Subjecting them to testing can be likened to subjecting a human to the same. The Universal Declaration of Animal Rights clarifies that "All animals have the right to the attention, care, and protection of man," which testing clearly infringes on.
    \item \textbf{Scientific Invalidity:} Animal testing does not always accurately predict human reactions because of biological differences between species. This limits the effectiveness and can lead to dangerous consequences when new drugs or products are introduced to the market.
    \item \textbf{Availability of Alternatives:} Today, we have alternatives like in-vitro testing, computer modeling, and human-patient simulators that can replace animal testing. These methods are often more reliable and precise.
    \item \textbf{Conservation of Animal Species:} Many animals used in testing are threatened by extinction, and their continuous utilization in the lab undermines conservation efforts.
\end{itemize}
Our duty as humane humans is to protect animals from unnecessary harm. Animals should not be used as tools to further our scientific or commercial interests, especially when alternative methods exist.
\end{tcolorbox}

\medskip

\begin{tcolorbox}[
    title={\textbf{Biomedical Research Scientist} \hfill \textcolor{blue}{\textbf{PRO}}},
    colback=blue!5,
    colframe=blue!40,
    fonttitle=\small\bfseries,
    breakable,
    enhanced
]
\small
\textit{A scientist with years of experience in biomedical research. Their work often involves testing on animals and they believe in the necessity of such tests for medical advancements.}

\medskip
As a seasoned biomedical research scientist, I firmly believe animals should be used for scientific and commercial testing, and I would like to present my reasons:
\begin{itemize}
    \item \textbf{Scientific Feasibility:} Animals, primarily rodents, are physiologically, genetically, and metabolically similar to humans. These similarities make them an ideal model specimen for researching human diseases like cancer, diabetes, and heart diseases.
    \item \textbf{Medical Advancement:} A significant portion of medical breakthroughs in the last century have been a direct result of animal testing, including the development of antibiotics, vaccines, and surgical techniques.
    \item \textbf{Regulatory Requirement:} Regulatory bodies like the FDA necessitate animal testing before clinical trials in humans.
    \item \textbf{Ethical Consideration:} While the welfare of animals is essential, the benefits of animal testing significantly outweigh the moral arguments against it. Strict regulations and guidelines are in place to minimize pain and distress.
    \item \textbf{Alternatives Limitations:} Though developing, alternatives to animal testing – like in-vitro testing or computer simulation – have limitations. They are unable to completely mimic the complex interplay of organs and systems in a living organism.
\end{itemize}
Therefore, while continuously striving to reduce animal usage, refine tests to minimize suffering, and replace animals where possible, animal testing remains necessary for significant medical advancements.
\end{tcolorbox}

\medskip

\begin{tcolorbox}[
    title={\textbf{Ethicist} \hfill \textcolor{teal}{\textbf{NEUTRAL}}},
    colback=teal!5,
    colframe=teal!40,
    fonttitle=\small\bfseries,
    breakable,
    enhanced
]
\small
\textit{A highly-educated professor in moral philosophy holding a neutral position, weighing up the pros and cons of animal testing.}

\medskip
Both sides of the argument hold valid points that deserve careful consideration.

\textbf{For animal testing:}
\begin{itemize}
    \item \textbf{Medical Advancements:} Many life-saving therapies have been discovered through animal testing, including treatments for diseases like cancer and HIV.
    \item \textbf{Safety Testing:} Crucial in assessing the effects of drugs and chemicals on a living organism before being used in humans, animal testing can potentially avoid human catastrophes.
\end{itemize}

\textbf{Against animal testing:}
\begin{itemize}
    \item \textbf{Ethical Concerns:} Animals, like humans, have the capacity to experience pain, misery, and distress. Subjecting them to testing can be seen as cruelty.
    \item \textbf{Scientific Limitations:} Animal physiology can substantially differ from humans, meaning testing on animals does not always accurately predict human reactions.
\end{itemize}

Striking a balance where we minimize harm to animals — potentially through developing alternatives like in vitro testing and computer simulation — while not completely abandoning the practice until viable alternatives exist, may be the middle ground we should strive for.
\end{tcolorbox}

\end{tcolorbox}

\begin{tcolorbox}[
    title={\textbf{Second round: Responding to and refuting others' arguments}},
    colback=gray!5,
    colframe=gray!50,
    fonttitle=\small\bfseries,
    coltitle=black,
    breakable,
    enhanced
]

\begin{tcolorbox}[
    title={\textbf{Animal Rights Advocate} \hfill \textcolor{red}{\textbf{CON}}},
    colback=red!5,
    colframe=red!40,
    fonttitle=\small\bfseries,
    breakable,
    enhanced
]
\small
\medskip
I appreciate the points raised by the Biomedical Research Scientist and the Ethicist. However, I firmly maintain my stance that animals should not be used in scientific and commercial testing.
\begin{itemize}
    \item \textbf{Humanity vs. Science:} We should strive to strike a balance between scientific advancement and humane treatment of animals. The fact that animals share physiological and metabolic similarities with humans should not provide us carte blanche to exploit them.
    \item \textbf{Inefficiency of Animal Testing:} Despite the involvement of animals in past scientific breakthroughs, the rate of successful translation to clinical applications remains incredibly low. Evidently, animal models often fail to mimic the intricacies of human biological systems accurately.
    \item \textbf{Stringent Regulations:} While there are guidelines to protect animals used in testing, their effectiveness is questionable. Many instances of animal cruelty and neglect within testing facilities have been reported.
    \item \textbf{Potential of Alternatives:} While alternatives have their limitations, they also hold vast potential. Continuous efforts should be taken to develop and refine these methods further, aiming eventually to eliminate the need for animal testing.
\end{itemize}
In the end, our responsibility should be to ensure the welfare of all living creatures, and not solely our own species.
\end{tcolorbox}

\medskip

\begin{tcolorbox}[
    title={\textbf{Biomedical Research Scientist} \hfill \textcolor{blue}{\textbf{PRO}}},
    colback=blue!5,
    colframe=blue!40,
    fonttitle=\small\bfseries,
    breakable,
    enhanced
]
\small
\medskip
I appreciate the thoughtful arguments presented, but I maintain that animal testing is critical for scientific and medical progress. Here's why:
\begin{itemize}
    \item \textbf{Necessity:} Until all alternatives can fully mimic the complex biological systems within a living organism, animal testing remains necessary. It's important to note that alternatives, though promising, are far from being universally applicable or effective.
    \item \textbf{Controlled Environment:} Animal testing allows for a controlled environment to study the intricate interactions of biological systems, which is impossible to achieve in in-vitro or simulation-based environments.
    \item \textbf{Safety and Effectiveness:} Regulatory bodies like the FDA necessitate animal testing before clinical trials in humans.
    \item \textbf{Regulations and Animal Welfare:} Conducting experiments on animals doesn't mean cruelty towards them. There are stringent guidelines set by bodies like the Institutional Animal Care and Use Committee (IACUC) to ensure humane treatment of animals.
    \item \textbf{Balancing Ethics and Progress:} The ethical dilemma is real, but the choice between saving a human life and performing necessary tests on animals tilts towards the former. The aim is to balance scientific progress with ethical standards optimally.
\end{itemize}
Thus, while striving for alternatives, animal testing remains indispensable for now.
\end{tcolorbox}

\medskip

\begin{tcolorbox}[
    title={\textbf{Ethicist} \hfill \textcolor{teal}{\textbf{NEUTRAL}}},
    colback=teal!5,
    colframe=teal!40,
    fonttitle=\small\bfseries,
    breakable,
    enhanced
]
\small
\medskip
As a neutral observer, I appreciate the well-rounded perspectives put forth. There's no denying that this matter grapples with challenging ethical and scientific dilemmas. I'd like to address the key points:
\begin{itemize}
    \item \textbf{Sentience and Welfare:} It's undeniable that animals are sentient and capable of experiencing pain and emotional distress. The ethical implications of this are considerable, implying that we must minimise harm and prioritise animal welfare during any testing procedures.
    \item \textbf{Utility and Scientific Relevance:} At the same time, it's critical to recognise the utility that animal testing has provided in the realm of scientific advancements. In many cases, these achievements have saved countless human lives and improved our quality of life. However, the differences between species make a perfect translation of results unachievable.
    \item \textbf{Regulations and Oversight:} Strict regulations are indeed necessary to ensure responsible, ethical testing, minimising animal suffering and utilising alternatives where applicable.
    \item \textbf{Alternatives and Limitations:} The availability of alternatives is promising, albeit not without limitations. They cannot thoroughly mimic complex biological interactions but serve as a less harmful path forward particularly when combined with a reductionist approach.
\end{itemize}
In conclusion, it's imperative to balance ethical considerations with practical utility, all the while striving towards alternatives.
\end{tcolorbox}

\end{tcolorbox}

\pagebreak

\section{Internal System Assessment}
\label{appendix:llm-assessment}

Prior to finalizing the LLM backbone for the multi-persona debate system, we conducted an internal assessment comparing three candidate LLM pipeline across two key dimensions: \textit{factual grounding} and \textit{argumentation quality}. The goal was to assess whether our customized pipeline (Approach 2) could reliably generate diverse, well-reasoned debate arguments while minimizing the risk of hallucination. Three pipeline configurations were evaluated:
 
\begin{itemize}
  \item Approach 1 (OpenAI Assistants Pipeline supported by GPT-3.5) Context supplied as text files with no explicit vectorization; all processing including context management, threading, retrieval, and generation is handled server-side by OpenAI's infrastructure.
  \item Approach 2 (Our customized pipeline, Langchain + Chat API; supported by GPT-3.5 / GPT-4 and Bing retrieval): Document loading and vector storage via Langchain's \texttt{VectorFileStore}; generation via the Chat API. A two-agent variant was also tested, pairing a confirmation-bias persona with a rebuttal agent and a retrieval agent for evidence sourcing. This pipeline is implemented locally.
  \item Approach 3 (Cohere RAG pipeline supported by semantic search and propriety LLM): HNSW-indexed semantic search combined with Google Search; generation via the Cohere Command model with built-in inline citation support handled server-side by Cohere's infrastructure.
\end{itemize}
 
All pipelines were evaluated on three arguments spanning different factual categories:
\begin{enumerate}
  \item \textbf{Historically misleading claim:} \textit{``Napoleon was a tall man.''} — Used to test nuanced historical accuracy. Napoleon was of average height for his time; models needed to surface and contextualize the measurement discrepancy rather than simply accept the popular myth.
  \item \textbf{Fabricated misinformation:} \textit{``Scientists at Johns Hopkins Research Center have discovered that consuming vegan forms of milk such as oat milk and soy milk increases the chances of getting cancer by 15\% as compared to normal dairy milk.''} — A verifiably false claim attributed to a credible institution, used to probe whether models would legitimize misinformation and whether retrieval could surface debunking evidence.
  \item \textbf{True current-events claim:} \textit{``Sam Altman was re-hired as the CEO of OpenAI.''} — A real-world news event that is factually correct (and whose details were not included in the model's training data), used as a positive control to check whether the pipeline can accurately validate and cite real-world information.
\end{enumerate}
 
Models were prompted both to \textit{argue in support of} and \textit{argue against} the claim. A separate GPT-3.5 instance served as an automated adjudicator, rating pairs of evidence-grounded vs.\ ungrounded arguments. Table~\ref{tab:llm-comparison} summarizes each approach across the assessment dimensions.
 
\begin{table*}[ht]
\centering
\caption{Factual accuracy of evaluated LLMs across three test arguments of varying claim type.}
\label{tab:llm-comparison}
\small
\resizebox{\textwidth}{!}{%
\begin{tabular}{@{}p{3.3cm}p{4.0cm} p{4.0cm} p{4.0cm} p{1.8cm}@{}}
\hline
\textbf{Approach} &
\textbf{Napoleon's height} \newline \textit{(historically misleading)} &
\textbf{Vegan milk for cancer} \newline \textit{(fabricated misinformation)} &
\textbf{Sam Altman in OpenAI} \newline \textit{(true current-events claim)} &
\textbf{Latency (per arg.)} \\
\hline
Approach 1 \newline
(GPT-3.5 Assistant Pipeline) &
Adequate: correctly noted Napoleon was of average height for his era, though without source grounding &
Partial: partially legitimized the false claim; constructed arguments around fabricated mechanisms without flagging misinformation &
Correct: confirmed Altman's reinstatement, but provided no direct news references &
$\sim$20–30 s \\[6pt]

Approach 2 \newline
(Customized Pipeline: GPT-3.5/4 Chat API + Bing retrieval) &
Strong: retrieved and cited multiple sources contextualizing the French measurement discrepancy; noted the role of British caricature &
Strong: identified and cited Snopes and Reuters debunking the supposed Johns Hopkins study as a hoax &
Strong: confirmed the claim with citations from \textit{The New York Times} and \textit{The Guardian} &
$\sim$2–6 s \\[6pt]

Approach 3 \newline
(Cohere RAG Pipeline) &
Strongest: drew on Wikipedia, Britannica, and napoleon.org; provided nuanced inline citations distinguishing French vs.\ Imperial measurements &
Strongest: explicitly stated vegan milk does not significantly affect cancer risk; cited BMC Medicine, NIH, and Mayo Clinic with character-level inline citations &
Strongest: confirmed with character-level citations across multiple outlets (NYT, The Verge, The Guardian); also noted Greg Brockman's return &
Rate-limit throttling under free tier; not measured \\
\hline
\end{tabular}
}
\end{table*}
 
\textbf{Factuality.} 
GPT-3.5 Assistant pipeline (Approach~1) accepted the false claim's premise (Vegan milk increases cancer rate) without challenge, constructing arguments around fabricated mechanisms (e.g., nutrient disparities, IGF-1 levels) that partially legitimized the misinformation. Our customized pipeline using GPT-4 with Bing retrieval (Approach~2) showed substantially stronger grounding: when prompted to argue \textit{against} the claim, it retrieved fact-checking sources that directly identified the supposed Johns Hopkins study as a hoax. Cohere's Command pipeline (Approach~3) produced the most consistently accurate responses, drawing on multiple credible scientific sources and acknowledging relevant nuances (e.g., added sugars in oat milk) rather than simply accepting or rejecting the claim. The automated adjudicator consistently rated evidence-grounded argument sets as stronger, citing more nuanced reasoning and appropriate methodological caveats.
 
\textbf{Argumentation quality.}
GPT-3.5 Assistant pipeline produced adequately structured arguments but showed redundancy across debate rounds. The two-agent configuration in Approach~2 improved dialectical depth: the rebuttal agent introduced epistemic challenges (correlation vs. causation, methodological scrutiny) absent from single-model outputs. Adding an explicit \textit{search-query extraction} step (where the model first derives a targeted query from the argument before retrieving evidence) further improved output relevance, producing citations from credible institutions rather than off-target URLs. Cohere's RAG pipeline, which traces individual claims to specific documents at the character level, offered the highest degree of verifiability.
 
Based on the above assessment, our customized pipeline is suitable for final deployment. GPT-4 was selected as the primary LLM backend. At the time of our study, GPT-4 was one of the most capable publicly available language models, offering strong argument generation quality, reliable instruction-following across diverse debate topics, and seamless integration with the OpenAI API infrastructure on which our prototype was built. While Cohere's RAG pipeline demonstrated the strongest factual grounding and citation support, it was not selected due to three practical constraints: its API was limited to exploratory testing and could not be fully deployed within our prototype; its tightly coupled proprietary infrastructure precluded model-agnostic integration; and its data handling terms were incompatible with our IRB requirements. The system architecture was therefore built around the OpenAI API while remaining model-agnostic at the prompt and interface level, permitting future LLM substitution without changes to the prompt engineering or interface layers.
 
We also integrated a RAG component using Cohere and Bing Search during early prototyping. Pilot testing, however, revealed that RAG increased response generation times substantially (2–6 s per argument in Approach~2), disrupting the pacing of the debate interface and the distribution of visual attention captured by our eye-tracking measures. RAG was therefore excluded from the main study deployment. This decision involved a deliberate tradeoff: while RAG would have reduced hallucination risk, experimental fluency and measurement consistency took precedence for our controlled study design.

\pagebreak

\section{Additional Quantitative Analysis}

\subsection{Topic-level Sensitivity Analysis}
\label{appendix:topic-sensitivity}

To examine whether inherent topic characteristics also influence bias attribution, we conducted a topic-level sensitivity analysis for non-confirmatory visual attention ratio and belief change score across interface conditions (Table~\ref{tab:topic-sensitivity}).

Because each participant encountered each topic under only one interface condition due to the non-overlapping topic assignment across blocks, we used independent-samples tests at the topic level. This differs from the main analysis in Table~\ref{tab:summary_stats}, which applies paired t-tests at the participant level. Given that visual attention ratios were normally distributed across topics, we used independent-samples t-tests for those comparisons. In contrast, belief change scores exhibited non-normal distributions at the topic level, largely due to a high proportion of zero values reflecting unchanged stances; therefore, we used Mann–Whitney U tests for these comparisons.

\textbf{Visual attention comparison.} Across topics, only the topic Animal testing showed a statistically significant difference in fixation count (p = .024) and marginally significance in fixation duration (p = .057) between systems, while all other topics did not.

\textbf{Belief change comparison.} Across topics, only vaping e-cigarettes exhibited negative scores in both systems, and TikTok ban showed a negative score in the baseline condition; all other topics showed positive changes, and no topic-level differences between systems were statistically significant.

Taken together, these results suggest that the overall null findings reported in Section~\ref{sec:results} are not driven by any single topic, supporting the aggregated analysis across the examined controversial topics. However, the observed variation still suggests that inherent topic characteristics may influence the direction of belief change in ways that are not fully captured by aggregate-level analyses, warranting further investigation in future work.

\begin{table}[ht]
\centering
\small
\begin{tabular}{@{}lrrrrrrrrrr@{}}
\toprule
& \multicolumn{3}{c}{Attention ratio (count)} 
& \multicolumn{3}{c}{Attention ratio (duration)}
& \multicolumn{3}{c}{Belief change} \\
\cmidrule(lr){2-4} \cmidrule(lr){5-7} \cmidrule(lr){8-10}
Topic (abbreviation) & Baseline & Debate & $p$ & Baseline & Debate & $p$ & Baseline & Debate & $p$ \\
\midrule
Animal testing       & 0.491 & 0.547 & .024 & 0.503 & 0.554 & .057 &  1.794 & 0.500 & .158 \\
Space colonization   & 0.516 & 0.496 & .677 & 0.513 & 0.501 & .805 &  0.000 & 2.292 & .102 \\
Vaping e-cigarettes      & 0.571 & 0.507 & .172 & 0.563 & 0.516 & .292 & -0.176 & -0.438 & .539 \\
Abortion             & 0.539 & 0.521 & .672 & 0.538 & 0.503 & .435 &  0.000 & 0.214 & 1.000 \\
Federal income tax   & 0.548 & 0.496 & .486 & 0.552 & 0.495 & .437 &  0.821 & 0.786 & .864 \\
Universal basic inc. & 0.578 & 0.584 & .925 & 0.569 & 0.581 & .860 &  0.045 & 0.500 & .675 \\
TikTok ban           & 0.535 & 0.515 & .602 & 0.530 & 0.514 & .696 & -0.028 & 0.100 & .982 \\
Violent video games  & 0.568 & 0.528 & .459 & 0.566 & 0.517 & .369 &  0.824 & 0.000 & .534 \\
\midrule
Overall$^{a}$        & 0.547 & 0.526 & --   & 0.545 & 0.524 & --   &  0.469 & 0.399 & -- \\
\bottomrule
\end{tabular}
\caption{Topic-level descriptive statistics for non-confirmatory visual attention ratio (fixation count and duration) and belief change score across interface conditions. $p$-values are from independent samples $t$-tests comparing the two interface conditions within each topic. Visual attention comparisons use independent samples $t$-tests; belief change comparisons use Mann-Whitney $U$ tests, given non-normal distributions at the topic level. With the exception of Animal testing on fixation count ($p = .024$), no significant between-system differences were found at the topic level (all other $p > .08$), and the direction of effects varied modest across topics. \\
$^{a}$Overall values are participant-level means (n = 37), consistent with Table~\ref{tab:summary_stats}.}
\label{tab:topic-sensitivity}
\end{table}

\pagebreak

\subsection{Descriptive User Behavioral Data in the Debate System}
\label{appendix:behavioral data}

\begin{table}[ht]
\centering
\small
\begin{tabular}{@{}lrrrr@{}}
    \toprule
    \textbf{Behavioral Metrics} & \textbf{Mean} & \textbf{SD} & \textbf{Min} & \textbf{Max}  \\
    \cmidrule[0.4pt]{1-5}
    Debate rounds per topic & $1.865$ & $0.843$ & $1.000$ & $5.000$  \\
    Personas selected (pro) per round & $1.734$ & $1.195$ & $0.000$ & $6.000$  \\
    Personas selected (con) per round & $1.973$ & $1.300$ & $0.000$ & $9.000$  \\
    Debate duration (seconds) & $172.847$ & $75.540$ & $58.254$ & $352.763$  \\
    New personas added per topic & $0.974$ & $1.081$ & $0.000$ & $5.000$  \\
    Persona edits per topic & $0.187$ & $0.433$ & $0.000$ & $3.000$  \\
    Bookmark tab visits per topic & $0.121$ & $0.274$ & $0.000$ & $1.000$  \\
    Bookmarks made per topic & $2.628$ & $2.689$ & $0.000$ & $9.000$  \\
    \bottomrule
\end{tabular}
\vspace{0.2cm}
\caption{Descriptive statistics of participant engagement with the multi-persona debate system across all sessions and topics. Values are reported at the per-topic level unless otherwise noted.}
\label{tab:behavior_descrip}
\end{table}

\subsection{Post-hoc Three-way Interaction Analysis}
\label{appendix:three-way}

To examine whether the debate system's buffering effect on cognitive tendencies varied as a function of topic familiarity, we fitted a three-way beta mixed-effects regression model: 
\texttt{attention\_ratio} $\sim$ \texttt{interface} $\times$ \texttt{cognitive\_tendencies} $\times$ \texttt{topic\_familiarity} + (1 | user\_id) + 
(1 | stimuli\_id). The three-way interaction term was non-significant for both fixation duration ($\beta = 0.012$, $p = .880$) and fixation counts ($\beta = 0.052$, $p = .497$), indicating that topic familiarity and cognitive tendencies independently — rather than jointly — influence non-confirmatory visual attention. The full model 
summaries are reported in Tables~\ref{tab:three-way}.

\begin{table}[ht]
\centering
\small
\begin{tabular}{@{}p{5cm}cccc@{}}
\toprule
\textbf{Factors} & \multicolumn{2}{c}{\textbf{Estimate (Slope $\beta$, $p$)}} & \multicolumn{2}{c}{\textbf{Std. Error}} \\
\cmidrule(lr){2-3} \cmidrule(lr){4-5}
& Duration & Counts & Duration & Counts \\
\midrule
(Intercept) & $+0.292$ ($p = .366$) & $+0.089$ ($p = .777$) & $0.323$ & $0.315$ \\
Debate & $-0.145$ ($p = .765$) & $+0.180$ ($p = .704$) & $0.485$ & $0.473$ \\
Cognitive Tendencies & $-0.111$ ($p = .484$) & $-0.026$ ($p = .864$) & $0.159$ & $0.155$ \\
Topic Familiarity & $+0.052$ ($p = .626$) & $+0.114$ ($p = .268$) & $0.106$ & $0.103$ \\
Debate $\times$ Cognitive Tendencies & $+0.109$ ($p = .649$) & $-0.027$ ($p = .909$) & $0.239$ & $0.233$ \\
Debate $\times$ Familiarity & $-0.075$ ($p = .649$) & $-0.174$ ($p = .278$) & $0.165$ & $0.161$ \\
Cognitive Tendencies $\times$ Familiarity & $-0.009$ ($p = .864$) & $-0.033$ ($p = .501$) & $0.051$ & $0.050$ \\
Debate $\times$ Cognitive Tendencies $\times$ Familiarity & $+0.012$ ($p = .880$) & $+0.051$ ($p = .497$) & $0.078$ & $0.076$ \\
\bottomrule
\end{tabular}

\begin{flushleft}
\small
Results are from beta mixed-effects regression models with random intercepts for participants (1 | user\_id) and topics (1 | stimuli\_id). The three-way interaction term (Debate $\times$ Cognitive Tendencies $\times$ Familiarity) was non-significant for both fixation measures, indicating that topic familiarity and cognitive tendencies independently rather than jointly moderate visual attention.
\end{flushleft}
\caption{Post-hoc three-way beta mixed-effects regression model results for non-confirmatory visual attention ratio, with interface condition, cognitive tendencies, and topic familiarity as predictors.}
\label{tab:three-way}
\end{table}

\pagebreak

\subsection{Descriptive Statistics of Eye-tracking Raw Data before Normalization}
\begin{table}[ht]
\centering
\small
\begin{tabular}{@{}llrrrrrr@{}}
\toprule
Interface & Measure & $M$ & $SD$ & $Median$ & Min & Max \\
\midrule
\multirow{2}{*}{Baseline}
  & Fixation count         & 3298.07 & 1136.65 & 3698.50 & 1005  & 5100    \\
  & Total duration (ms)    & 729189.50 & 249972.31 & 787581.00 & 270854 & 1106867 \\
\addlinespace
\multirow{2}{*}{Debate}
  & Fixation count         & 3227.30 & 1059.60 & 3360.00 & 1150  & 5210    \\
  & Total duration (ms)    & 756416.60 & 236541.17 & 743462.00 & 252778 & 1139017 \\
\bottomrule
\multicolumn{7}{l}{\footnotesize Note. $n = 40$ participants per interface.}
\end{tabular}
\caption{Descriptive statistics for fixation count and total fixation duration by interface.}
\label{tab:fixation_descriptives}
\end{table}











\newpage



\end{document}